\documentclass[journal,twoside]{ieeecolor}
\usepackage{cite}
\usepackage{tmi}
\usepackage{amsmath,amssymb,amsfonts}
\usepackage{algorithmic}
\usepackage{graphicx}
\usepackage{textcomp}
\usepackage{caption}
\usepackage{multirow}
\usepackage{float}
\usepackage{breqn}
\usepackage{subcaption}

\begin{document}

\title{\textbf{SAFRON: Stitching Across the Frontier for Generating Colorectal Cancer Histology Images}}

\author{Srijay Deshpande, Fayyaz Minhas, \\ Simon Graham, Nasir Rajpoot
\thanks{Srijay Deshpande, Fayyaz Minhas, Simon Graham and Nasir Rajpoot are with the Department of Computer Science, University of Warwick, Coventry, UK. Nasir Rajpoot is also with the Department of Pathology, University Hospitals Coventry and Warwickshire, UK and The Alan Turing Institute, London, UK. \\
Corresponding author email: srijay.deshpande@warwick.ac.uk.}}

\maketitle

\begin{abstract}

Synthetic images can be used for the development and evaluation of deep learning algorithms in the context of limited availability of data. In the field of computational pathology, where histology images are large in size and visual context is crucial, synthesis of large high-resolution  images via generative modeling is a challenging task. This is due to memory and computational constraints hindering the generation of large images. To address this challenge, we propose a novel SAFRON (Stitching Across the FRONtiers) framework to construct realistic, large high resolution tissue image tiles from ground truth annotations while preserving morphological features and with minimal boundary artifacts. We show that the proposed method can generate realistic image tiles of arbitrarily large size after training it on relatively small image patches. We demonstrate that our model can generate high quality images, both visually and in terms of the Frechet Inception Distance. Compared to other existing approaches, our framework is efficient in terms of the memory requirements for training and also in terms of the number of computations to construct a large high-resolution image. We also show that training on synthetic data generated by SAFRON can significantly boost the performance of a state-of-the-art algorithm for gland segmentation in colorectal cancer histology images.
Sample high resolution images generated using SAFRON are available at the URL: https://warwick.ac.uk/TIALab/SAFRON

\end{abstract}



\begin{figure*}[!hbtp]
\centering
\includegraphics[width=400pt,height=140pt]{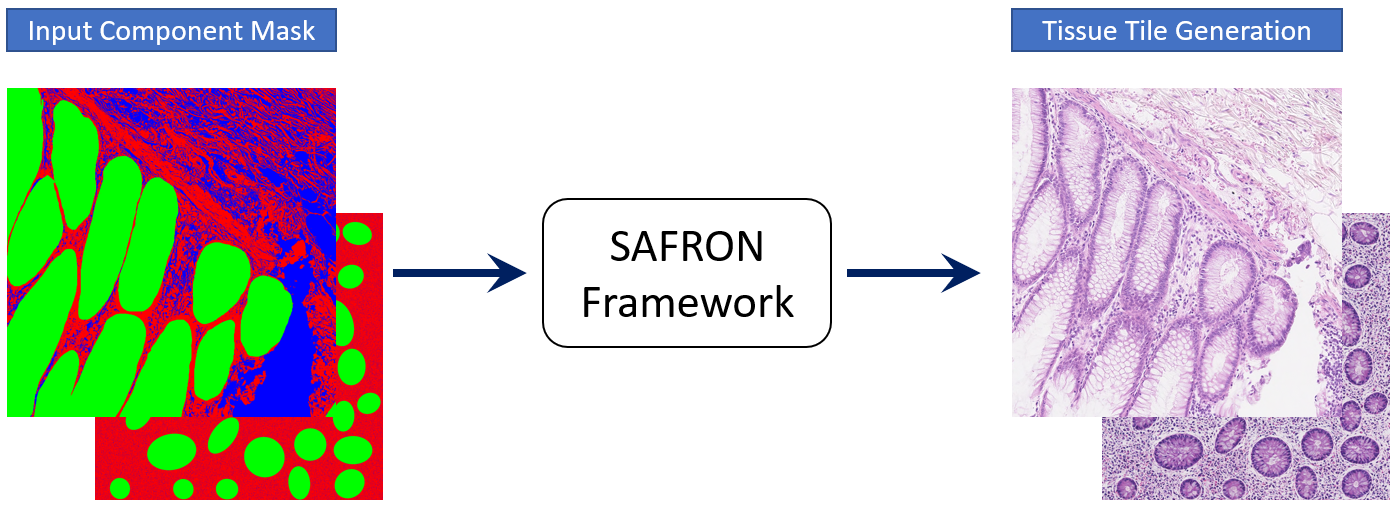}
\caption{A high-level concept diagram of SAFRON: Images on the left show tissue component masks that are obtained either from available datasets or constructed using a generative model described in section \ref{gettingcomponentmasks}. The SAFRON framework takes the masks as its input and constructs the synthetic tissue tiles (shown on the right) after stitching smaller synthetic sub-tiles generated using a GAN model, which is part of the SAFRON framework (see Figure \ref{safron_architecture}). The synthetic image data and the corresponding annotation masks can be used for training and evaluation of various algorithms such as gland segmentation.} 
\label{safron_concept}
\end{figure*}

\section{Introduction}

Automated analysis of histology whole-slide images (WSIs) has received a great deal of attention in the field of medical image analysis in recent years. Several deep learning approaches for problems in the area of computational pathology have been proposed. These include tumor segmentation (\cite{tumorsegment1,bejnordi2017diagnostic,qaiser2019fast}), segmentation of glands (\cite{graham2019mild, chen2017dcan}), cancer grading (\cite{cgrading1,shaban2020context,zhou2019cgc}) and nuclei detection and classification (\cite{ncd1,graham2018sams,graham2019hover,sirinukunwattana2016locality}). Digital WSIs are typically quite large in size, potentially containing several billions of pixels at the highest magnification. Therefore, development of efficient machine learning algorithms for histology slides remains a major challenge, due to limitations in processing capacity and memory storage. 

Deep learning models are \textit{data hungry} in nature, requiring large and high quality annotated datasets. This problem is exacerbated in medical image analysis, where data annotation needs to be done by an expert clinical specialist. To overcome the difficulty of curating large datasets, generative modelling of synthetic images has become an active area of research in recent years and has resulted in the development of numerous models (\cite{senaras2018optimized,senaras2018creating,quiros2019pathology}) for synthetic tissue image generation in the area of computational pathology (CPath). Large high resolution image tiles are useful because they provide additional context, which can assist the diagnosis in computational pathology (\cite{shaban2020context,bejnordi2017context}). However, synthetic generation of such large images is challenging due to their demands on memory and processing capacity. Synthetic images can also be useful where legal and ethical barriers prohibit sharing of genetic and image hampering the development of algorithms for tasks like detection of Microsatellite Instability (MSI) and other bio-markers. Recently, \cite{jacobkathermsi} developed the Histology CGAN to generate histology images for MSI detection.

In this paper, we propose the SAFRON (Stitching Across the FRONtier) framework for generating synthetic, annotated high quality histology images. The concept diagram of our proposed framework is illustrated in Figure \ref{safron_concept}. Specifically, the framework generates large dimensional colorectal tissue images based on input tissue component masks by methodically stitching together generated tissue regions within a large tissue tile. In order to generate realistic images with a high perceptual quality, our framework utilises adversarial training. To the best of our knowledge, this is the first framework that can generate histology images of large arbitrary sizes. This paper extends an earlier version (\cite{sashimi}) with the following major contributions:

\begin{enumerate}

\item We propose a computationally cheap and memory efficient framework that can generate tissue tiles of much larger sizes than the ones used for training. The SAFRON framework shows the ability of seamless generation of large tiles preserving homogeneity and edge crossing continuities between the adjacent patches. 

\item We demonstrate SAFRON's better computational efficiency compared to other existing methods for generating large high resolution images, in terms of memory demand for training the network and number of computations required to construct an image.
 
\item We show that SAFRON can generate high dimensional synthetic colorectal tissue tiles of dimensions greater than $7000\times9000$ pixels. As far as we are aware, it is the first framework that can generate such large arbitrary high resolution tissue images.

\item Our proposed framework has the ability to generate an unlimited amount of annotated synthetic data. As an example, we show that augmenting training data for gland segmentation with the synthetic data generated by SAFRON shows significant improvement in terms of segmentation accuracy.  

\item Our qualitative and quantitative evaluation on the CRAG (\cite{graham2019mild}) and DigestPath (\cite{digestpath}) datasets shows that the synthetic images are constructed while preserving morphological characteristics and can be useful for training and evaluation of gland segmentation algorithms. We also perform an ablation study to validate the importance of various components and design strategy of the SAFRON framework.

\end{enumerate}

The remainder of this paper is organised as follows. In the subsequent section, we go through the related work on relevant methods of image generation. In section {\color{red}{\ref{materials_methods}}}, we present the details of the SAFRON framework along with a detailed description of its components. In Section {\color{red}{\ref{exp_results}}}, we perform a detailed evaluation of the proposed SAFRON framework on the CRAG (\cite{graham2019mild}) and DigestPath (\cite{digestpath}) datasets. Later, in section {\color{red}{\ref{complexityanalysis}}}, we show its computational efficiency compared to other existing frameworks for generation of high resolution images. Finally, we conclude with future directions for this work.

\begin{figure*}[!hbpt]
\centering
\includegraphics[width=350pt,height=220pt]{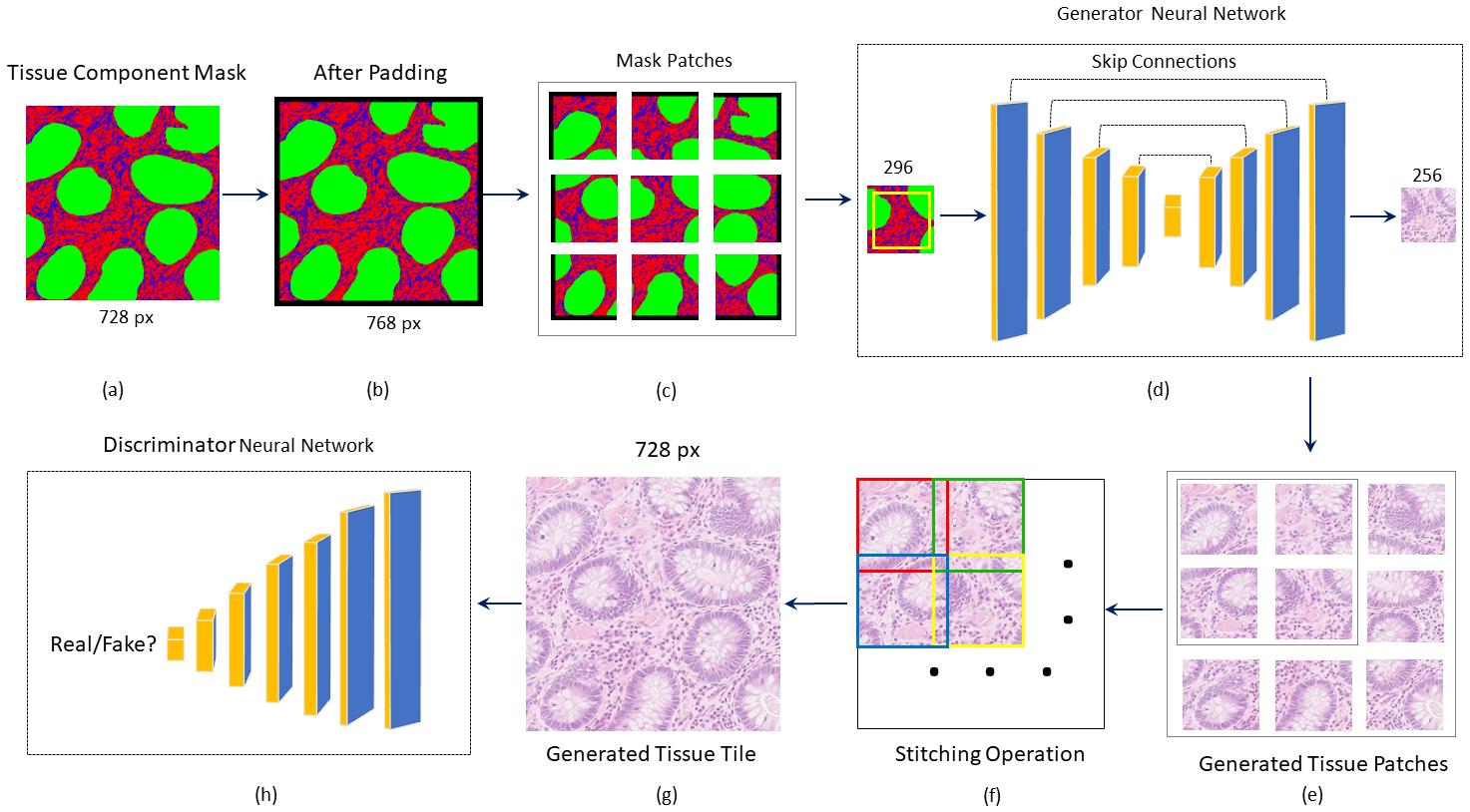}
\caption{An overview of the proposed SAFRON framework for training phase of high resolution tissue image generation. Tissue component mask (a) of size $728\times728$ is first divided into patches (c) after zero padding on all sides (b). We apply padding to ensure the dimensions of the generated tissue tile is the same as the dimensions of the input tissue component mask. Each patch of size $296\times296$ is passed through the generator neural network (d) creating corresponding tissue image patch of resolution $256\times256$. The generated tissue patch corresponds to the yellow colored square box shown in component mask patch (in (d)). Generated tissue patches (e) are then stitched (f) constructing the final tile (g) which is then passed entirely through the discriminator neural network (h) which predicts its realism. The pixels in the overlapped region (in (f)) are computed by spatial averaging of overlapping pixels. During inference time, as the patch-level generator is responsible for generating patches of tissue tiles from corresponding input mask patches and stitching of adjacent patches is independent of tile size, the proposed framework can be used to construct large histology images of arbitrary sizes.}
\label{safron_architecture}
\end{figure*}

\section{Related Work}

\subsection{Generative modelling of tissue images}

In earlier trends, classical machine learning was applied widely to model various tissue components and simulating tissue microenvironment. For example, SimuCell toolbox (\cite{simucell}) managed to simulate heterogeneous cell populations expressing various protein markers. \cite{zhao2007automated} proposed a machine learning based approach to generate realistic cells with labelled nuclei, membranes and a protein expressed in a cell organelle. \cite{kovacheva2016model} came up with a model named TheCOT to simulate colorectal cancer tissue images with tumor microenvironment. It allowed user to change appearance of tissues using user-based parameters. Though these models showed ability to model tissue components at a granular level, they were lacking in attaining realism in generated images.  

After the adoption of deep learning methods in generative modeling of images, the ability of generative adversarial networks, GANs (\cite{goodfellow2014generative}) to use adversarial training to generate realistic natural images with high perceptual quality motivated researchers to adopt them for synthesising histology images. For instance, \cite{quiros2019pathology} presented Pathology-GAN to generate high fidelity cancer tissue images. The model showed to capture clinically or pathologically meaningful representations within cancer tissue images and also demonstrated the induction of ordered latent space based on tissue characteristics. It allowed to perform linear arithmetic operations to change the high-level tissue image. \cite{medicalgan} delivered an adversarial learning-based approach to generate medical images for the task of medical image tissue recognition. \cite{robustlabelsynthesize} proposed an unsupervised pipeline to construct both histopathology tissue images and their corresponding nuclei masks to train the nuclei segmentation supervised algorithm. The model computed the importance weights of training samples using output from its discriminator network to train the task-specific nuclei segmentation algorithm. The conditional GANs (\cite{mirza2014conditional}), cGANs inspired researchers to adopt them for tissue image synthesis as they showed ability to synthesise high fidelity realistic images conditioned on the known ground truth. \cite{senaras2018optimized} proposed a cGAN based model to construct breast cancer tissue images from nuclei segmentation masks. Though these methods showed an ability to generate realistic tissue images, they are unable to produce large dimensional images of high resolution due to limitations in processing power and available memory. Collectively, these methods require rescaling histopathology images for implementation, leading to loss of high-resolution information about tissue components. 

\subsection{High resolution image generation}

The large dimensional image tiles of high resolution provide richer and high-quality context for designing algorithms, which is important for diagnosis in computational pathology \cite{shaban2020context,bejnordi2017context}. Despite the indisputable success in generating realistic images, generative adversarial networks (GANs) (\cite{goodfellow2014generative}) still struggle to generate high resolution images. One reason is the real images are easier to distinguish from their generated counterparts at high resolutions which hinders the training of GANs (\cite{progan}). To deal with it, \cite{progan} proposed ProGAN which adopted the generating the low resolution image initially and add finer details to it by expanding the network layer-wise throughout the training. Similarly, the SR-GAN \cite{srgan} employed the network on deep residual connection (ResNet) that super-resolved the photo-realistic images with a 4× up-scaling factor. Though these methods successfully showed the generation of high-quality images, they still require huge amount of memory to train networks on very large images (of dimensions more than $1024\times1024$ pixels), which means they require special and expensive hardware for image generation. This memory requirement increases exponentially with the image size which makes training these networks practically infeasible. To overcome this challenge, \cite{multiscalegan} presented a memory efficient Multi-scale GAN framework to construct high resolution tomography/MRI images. They considered the approach of learning a low-resolution image first, and then
generating patches of consecutively high resolutions using super resolution GANs conditioned on previous scales. However, as the network gets expanded with the target image resolution, it becomes computationally expensive and incurs high image generation latency. One unavoidable disadvantage of all of the above mentioned generative models is that they need a training data having large number of images of the same target resolution to train the model, which is difficult to obtain for very high resolution images.

\section{Materials and Methods}
\label{materials_methods}

The construction of large-size synthetic images is computationally demanding due to the high computational and memory requirements of image generation neural networks such as variational auto-encoders \cite{kingma2019introduction} and standard generative adversarial networks \cite{goodfellow2014generative}.  
In this work, we propose a novel method which can generate large high resolution tissue images of arbitrary sizes conditioned on input tissue component masks. The proposed method, called Stitching Across the FRONtier (SAFRON), aims to provide annotated synthetic data which can then be used for the training and evaluation of computational histopathology algorithms. The proposed framework attempts to generate small tissue image patches based on local tissue components such as glands, stroma and background specified as an input mask and methodically stitches the generated patches in a seamless manner to construct an entire tissue tile, thus, overcoming computational limitations.

\subsection{Datasets}

For training and performance evaluation of the proposed SAFRON framework, we require annotated real image data in terms of tissue component specification for conditioning the generation of synthetic images. In this work, we consider two datasets for our experiments: CRAG\footnote{ https://warwick.ac.uk/TIA/data/mildnet/} (\cite{graham2019mild,awan2017glandular}) and Digestpath \cite{digestpath}. 

The CRAG dataset has been widely used for the development of numerous gland segmentation methods (\cite{graham2019mild, graham2020dense, chen2017dcan, ding2020multi}). It (\cite{graham2019mild}\cite{awan2017glandular}) consists of 213 colorectal histopathology cancerous tissue images with variable cancer grades and a size of $1512\times1516$ pixels at a resolution of 0.55$\mu$m/pixel (20$\times$ objective magnification). Each image in the dataset belongs to one of three classes based on structure and morphology of glands: normal/healthy (no deformation of glands), low-grade cancer (slightly deformed glands), and high-grade cancer (highly deformed glands). Each image in the dataset is also associated with a gland segmentation mask which specifies glandular, non-glandular (stromal tissue) and background (based on raw pixel thresholding) regions. The input mask can be seen on the left part of Fig. \ref{safron_concept}, where the glands are shown in green, the stroma in red and the background in blue. This input segmentation mask is used as a tissue component mask to condition the generation of synthetic images through the proposed framework. In this paper, we have focused on generation of images with normal grades only. The dataset contains 48 large colon tissue cancerous images with normal grade. of which we use 39 (the CRAG train set) to extract training data and 9 (CRAG test set) to extract testing data.

We also use the DigestPath (\cite{digestpath}) dataset to assess the performance of our algorithm. The dataset is collected from the DigestPath2019 challenge\footnote{https://digestpath2019.grand-challenge.org/}. Similar to the CRAG dataset, it also contains tissue images with corresponding glandular masks. However, the image tiles are generally larger with an average size of $5000\times5000$ pixels. This dataset originally contains annotations for malignant lesions only. In order to obtain a tissue segmentation mask that shows glandular, stromal and background regions, we used a semi-automatic approach. For this purpose, we first trained trained a gland segmentation model (\cite{graham2019mild}) on the GlaS (\cite{glas1,glas2}) and CRAG (\cite{graham2019mild}) datasets and obtained gland segmentation masks for images in the DigestPath dataset which were manually refined. We have used a total of 46 large images from this dataset for our experiments: 20 images (DigestPath train set) to extract training data and 26 images (DigestPath test set) for extraction of testing data.

\subsection{Architecture}

The proposed framework is based on generative adversarial neural networks. A conventional generative adversarial network (GAN) (\cite{goodfellow2014generative}) consists of two components, a generator and a discriminator, which are trained simultaneously such that the generator learns to produce realistic images from an underlying image distribution whereas the discriminator attempts to discriminate between real and generated images. On successful training on real images, the generator is used for generating realistic images which are not part of the training set. Conditional generative adversarial networks (\cite{mirza2014conditional}) are capable of generating images controlled by some ground-truth input. 

\begin{figure*}[hbt!]
\centering
\includegraphics[width=420pt,height=120pt]{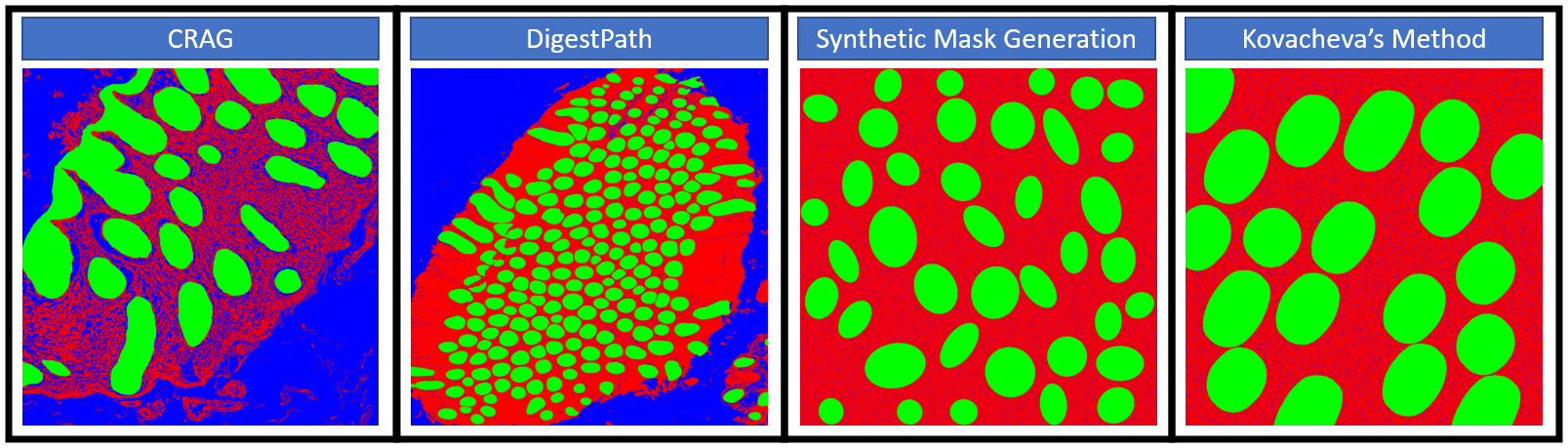}
\caption{Variety of tissue component masks used in this work to generate realistic tissue tiles. From left, first two component masks are collected from existing datasets CRAG and DigestPath, third component mask is generated using synthetic mask generation method, whereas the fourth one is constructed using \cite{kovacheva2016model}}
\label{random_tissue_gen}
\end{figure*}

Figure \ref{safron_architecture} presents an overview of the proposed framework. The input to the framework is a tissue component mask which specifies different tissue regions to be generated by the proposed network. Similar to existing methods for histology image generation based on conditional adversarial networks, the proposed framework also uses a generator network to generate small tissue image patches from a patch of the input tissue component mask. The patches are then then stitched to generate large image tiles. However, the major challenge here is to preserve global coherence and edge crossing continuity between adjacent patches so that the generated tile appears homogeneous and seamless while staying within image size limitations imposed by memory and computational processing constraints. In SAFRON, the size of the input tissue component mask patch is kept larger ($296\times296$ pixels) than that of the output tissue image patch ($256\times256$ pixels) to encourage context aware generation without seams at patch borders. To enable this, we apply zero padding to the original tissue component mask, so that after stitching together the generated patch-level results, the final tile will have the same dimensions as the input tissue component mask. Furthermore, in contrast to existing methods in which both  generator and discriminator networks work at patch level, the generator network in SAFRON generates ($256\times256$ pixels) patches whereas the discriminator network is trained over ($728\times728$ pixels) tile-level images \emph{after} stitching. Since the SAFRON discriminator treats the generated tile as a single large image, this enforces global coherence with minimal computational overhead. Additionally, adjacent patches are constructed with a small overlap which helps to reduce boundary artifacts. Furthermore, since the generator generates small ($256\times256$ pixels) image patches, it has a low memory and computational processing fingerprint.

Since the patch-level generator is responsible for generating patches of tissue tiles from corresponding input mask patches and  stitching of adjacent patches is independent of tile size, the proposed framework can be used to construct large histology images of arbitrary sizes. Intuitively, the framework learns to generate local regions and stitch them in a way that the generated tile exhibits a seamless appearance without boundary artifacts between adjacent regions. Below, we discuss different components of the proposed framework in detail. 

\subsection{Acquisition of Tissue Components Masks}
\label{gettingcomponentmasks}

A tissue component mask is used to specify different tissue regions as input to the proposed framework. For training and performance evaluation, we use tissue segmentation masks in CRAG and DigestPath datasets as discussed in the Dataset section above. However, for some experiments, we use two additional approaches to construct tissue component masks: i) Synthetic mask generation, and, ii) THeCoT Spatial Model of Tumour Heterogeneity in Colorectal Adenocarcinoma Tissue by \cite{kovacheva2016model}.

For synthetic mask generation, we first place elliptical objects of variable sizes as glands on a blank tissue component mask of desired size. A total of 3 to 7 such objects are added at random spatial locations in a ($100\times100$) pixel area. These `glandular' objects are colored in green to discriminate them from stromal (`red') and background (`blue') regions which are added in the remaining area with random color filling using a binomial distribution with probabilities of 0.9 and 0.1, respectively.

THeCoT constructs synthetic images along with their gland segmentation mask based on input parameters like cancer grade, cellularity, cell overlap ratio, image resolution and objective level. The gland segmentation masks generated by this model can be used as tissue component masks in SAFRON with stromal and background regions added in a similar manner as used for synthetic component mask generation. In comparison to the first method, this allows us to generate tissue component masks in a parameter-controlled manner.

Sample tissue component masks obtained from these approaches are shown in Figure \ref{random_tissue_gen}.

\subsection{Patch Generation}

We denote a given input component mask and corresponding real histology image as $X$ and $Y$, respectively, which can be modelled as ordered sets of patches, i.e., $X = \{ x_{r,c}\}$ and $Y=\{y_{r,c}\}$ with each patch parameterized by its center grid coordinates $(r,c)$ in the corresponding image. It is important to note that the size of an input component mask patch is kept larger ($296\times296$ pixels) than the size of tissue patch ($256\times256$ pixels) to encourage context aware image generation. Furthermore, patches are generated with an overlap of 20 pixels between adjacent patches which allows for seamless stitching.  

\begin{figure*}[hbtp]
\centering
\includegraphics[width=0.9\textwidth]{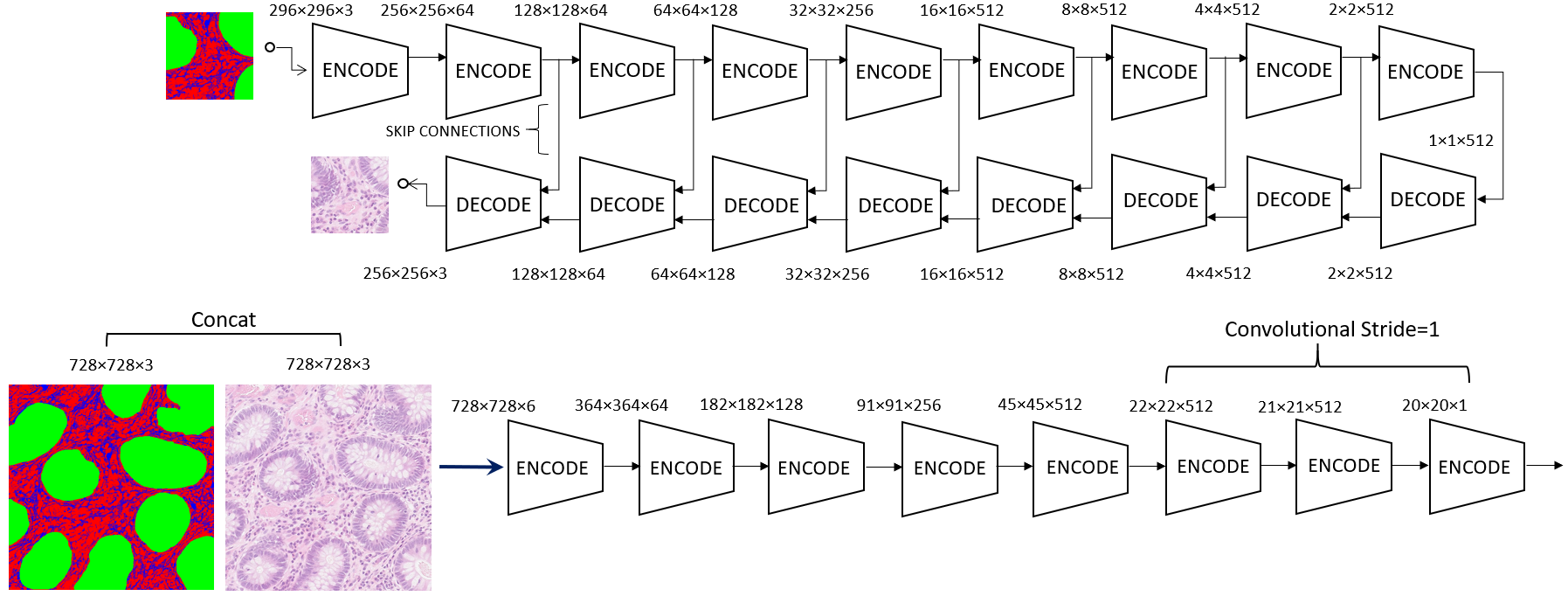}
\caption{Architectures of the generator (above) and discriminator (below). We train the generator at patch-level with input size $296\times296$ and output size $256\times256$, and train the discriminator at tile-level with input as concatenation of mask and image of sizes $728\times728$.}
\label{gen_discrim}
\end{figure*}

\begin{figure}[hbt!]
\centering
\includegraphics[width=220pt,height=82pt]{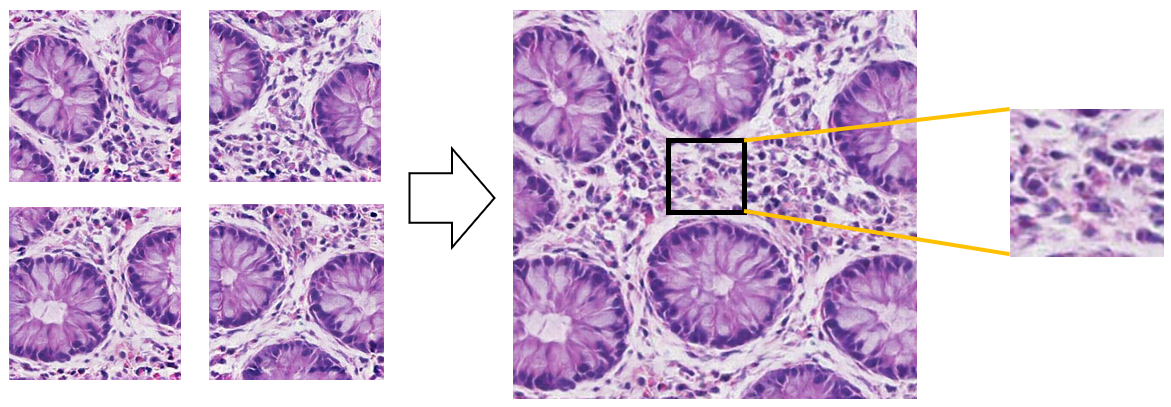}
\caption{Stitching operation: four spatially adjacent constructed patches from the generator are stitched (averaging the overlapped portion), combined region is shown in the middle figure which looks homogeneous and seamless. The merged portion at the middle of region is zoomed and shown in the rightmost figure}
\label{stitching}
\end{figure}

The proposed framework generates a tissue image patch $y'$ from a corresponding tissue component mask patch $x$ using a generator neural network $G$, i.e., $y'=G(x;\theta_G)$, where $\theta_G$ denotes the trainable weights of the generator.

\subsection{Stitching}

The patches generated by the SAFRON generator network $Y'=\{y'_{r,c}\}$ for a given input tissue component mask are stitched based on the spatial coordinates of their corresponding tissue component mask patches. Specifically, pixels in overlapping regions between adjacent patches are spatially averaged in the stitching process. The results of the stitching operation on one of the test image is shown in figure \ref{stitching}. It clearly shows that the stitched image does not have any seam or block artifacts after training.

The major challenge while generating the tissue tile, is to preserve the global coherence across it. For this purpose, the whole generated tile $Y'$ is consumed by the discriminator network $D(X,Y';\theta_D)$ with trainable weight parameters $\theta_D$.

\subsection{Generator and Discriminator Networks}

To construct an image patch from relatively bigger region of tissue component mask, we design an U-Net (\cite{ronneberger2015u}) based generator for our framework. The generator, $G(x;\theta_G)$ has a U-shaped structure with two components: encoder and decoder. The architecture of the generator is shown in Fig. \ref{gen_discrim}. The encoder takes a tissue component mask patch as input and constructs a low dimensional representation. The decoder then generates a tissue image patch corresponding to the input component mask patch based on encoder outputs. 

The encoder is made up of a series of encoding blocks, where each block performs the following operation: pass the input through a convolutional layer followed by batch normalization and finally output after applying leaky-relu activation. The decoder consists of a series of decoding blocks: each block up-samples the input to a higher dimension by first passing the input through a deconvolutional layer, followed by batch normalization and leaky-relu activation. The encoder and decoder operations can be visualised in section 7.1 (in supplementary materials). Similar to the generator used in \cite{isola2017image}, the generator utilises skip-connections between symmetric encoding and decoding blocks in the generator. Skip-connections give the generator flexibility to bypass the encoding part to subsequent layers and enable consideration of low level features from earlier encoding blocks in the generator.

The architecture of the discriminator used in this work is shown in figure \ref{gen_discrim}. The discriminator assigns a degree of realism to a generated or a real tile through its association with the corresponding tissue component mask. The discriminator network is inspired from the PatchGAN discriminator (\cite{isola2017image}). As shown in figure \ref{gen_discrim}, it has a series of encoding blocks which take a ($728\times728$) image tile and the corresponding mask as input and produce a $20\times20$ matrix that signifies the degree of realism of corresponding regions in the image tile. These elements of this $20\times20$ matrix are averaged to generate a single decision score from the discriminator.

\subsection{Adversarial Training and Inference}

For training, we use tissue component masks and their corresponding real image tiles of size ($728\times728$) which are obtained from images in a training dataset. The generator and discriminator networks are trained in an adversarial manner such that the generator tries to generate realistic images whereas the discriminator aims to determine if an image is real or generated from the generator. The generator $G(x;\theta_G)$  uses ($296\times296$)-sized tissue component mask patches $x\in X$ to generate corresponding ($256\times256$) patches $y'=G(x;\theta_G)$ of the tissue image which are stitched to generate a ($728\times728$) tile image. Without introducing further notation, we denote the stitched tile image corresponding to an input tissue component mask $X$ as $Y'=G(X;\theta_G)$. The discriminator network $D(X,Z;\theta_D)$ produces a realism score for a tile $Z$ which then used in Adversarial training. Adversarial training ensures that the generator learns to generate seamless images with global morphological coherence. 

The complete framework with trainable parameters $\{\theta_G,\theta_D\}$ is trained by minimizing a loss function with the following two components:

\textbf{Reconstruction Loss:} This loss component captures tile reconstruction error between original and generated tiles after stitching. Specifically, we use the expected value of the $L_1$ loss between actual and generated tile in the training dataset as given as below : 

\begin{equation}
   L_R(\theta_G)= E_{X,Y \sim p_{data}(X,Y)}\left \|Y - G(X;\theta_G)  \right \|_1
\label{l1}
\end{equation}

\textbf{Adversarial Loss:} The generator and discriminator networks are trained in an adversarial manner with the a cross-entropy based loss function which forces the generator to generator realistic images and the discriminator to classify between real and generate images: 

\begin{dmath}
L_{adv}(\theta_G,\theta_D) = E_{X,Y \sim p_{data}(X,Y)}[log D(X,Y;\theta_D)] + E_{X \sim p_{X}(X)}[log(1-D(X,G(X,\theta_G);\theta_D)]
\label{advloss} 
\end{dmath}

The overall learning problem can then be expressed as a the following adversarial optimization problem based on the linear combination of adversarial and reconstruction losses:

\begin{equation}
    \min_{\theta_G} \max_{\theta_D} \lambda_R L_R(\theta_G) + \lambda_{adv} L_{adv}(\theta_G,\theta_D)
\label{loss}
\end{equation}

The hyper-parameters, $\lambda_R$ and $\lambda_{adv}$, control the relative contributions and scaling of reconstruction and adversarial losses and their values  are set to 1 and 100, respectively, through cross-validation based tuning.

It is important to note that the proposed framework is trained to generate globally consistent patches given an input tissue component mask. Consequently, at inference time, we can use the trained SAFRON generator to effectively generate a large image of arbitrary size as the stitching mechanism is independent of the tile size that was used for training.

\begin{figure*}
\centering
\centering
 \makebox[\textwidth]{\includegraphics[width=\textwidth]{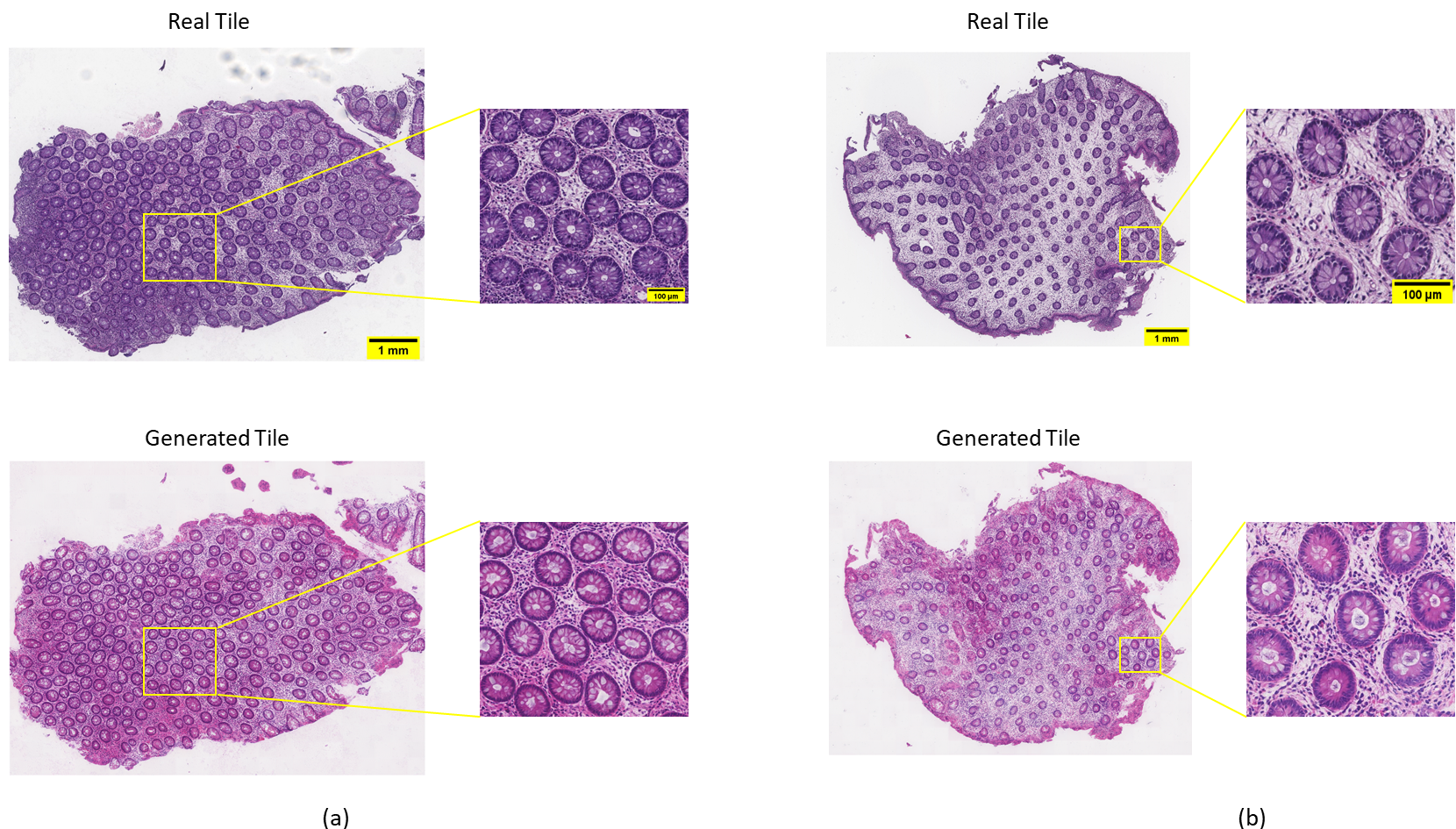}}
\caption{Constructed tiles along with their real versions from DigestPath dataset with sizes: (a) $5914\times4475$ pixels and (b) $6526\times5268$ pixels}
\label{image_results_digestpath}
\end{figure*}

\section{Experiments and Results}
\label{exp_results}

In this section, we present visual and quantitative results of the quality of generated images. We demonstrate how synthetic images can be used for training and performance assessment of gland segmentation algorithms. We show how SAFRON generated data can be used to augment limited training datasets and improve data efficiency of gland segmentation methods. Finally we demonstrate the importance of SAFRON design using ablation studies. 

\subsection{Experimental Settings}
\label{results_exp_settings}

To train the SAFRON framework, we extract 975 square tiles of size $728\times728$ with a stride of 200 pixels from the CRAG train set (defined in section 3) and 1,150 images of same size and 512 stride from DigestPath train set. For the purpose of testing, we extract images of various resolutions from the CRAG test and DigestPath test set. The network is trained for 100 epochs with Adam optimization and an initial learning rate $10^{-4}$, initial momentum 0.5 and batch size of one tile with multiple patches.

The Multi-Scale GAN (\cite{multiscalegan}) framework that generates high resolution medical images, is chosen as a baseline model for comparative analysis. The Multi-Scale GAN first generates a low resolution image and later increases its resolution by generating high resolution patches depending upon the previous scale.  In order to utilize this model in our use case, we employed Pix2Pix (\cite{pix2pix}) to generate tissue images of dimensions $256\times256$ from the tissue component mask and later employed a series of SR-GANs (\cite{srgan}) to increase its resolution depending on the previous scale. Unlike our framework, which trains on relatively smaller images than target images, Multi-scale GAN requires a set of images of same resolution it targets to generate. A large set of such annotated high resolution images is not easy to acquire. We extracted 141 images of resolution $4096\times4096$ with a stride of 500 from DigestPath train set for training the Multi-scale GAN, and 752 square tiles of resolution $2048\times2048$ with the same stride. To train the Multi-scale GAN on $4096\times4096$ images, we trained the pix2pix network in its first layer and it mapped tissue component mask of size $256\times256$ to tissue image of the same size. Later, we used multiple iterations of SR-GAN to generate an image of size $4096\times4096$ or $2048\times2048$ as needed for comparison. Unlike in our case, the Multi-scale GAN architecture needs to be altered depending upon the target image resolution.

We also considered the other state-of-the-art models for high-quality image generation such as Pix2pix (\cite{pix2pix}), SR-GAN (\cite{srgan}) and ProGAN (\cite{progan}). However, they can not be adapted for generation of high dimensional digital pathology images in our system setup due to memory and computational constraints. The detailed comparative analysis of memory requirements and runtime computations between different existing frameworks for high resolution image generation is given in section \ref{complexityanalysis}. All frameworks are implemented using Tensorflow (\cite{tensorflow})/Keras framework (\cite{keras}), on NVIDIA GeForce GTX TITAN X single GPU system with a dedicated RAM of 12GB.

\subsection{Visual Assessment}

\begin{figure*}[htb!]
\centering
 \makebox[\textwidth]{\includegraphics[width=400pt,height=150pt]{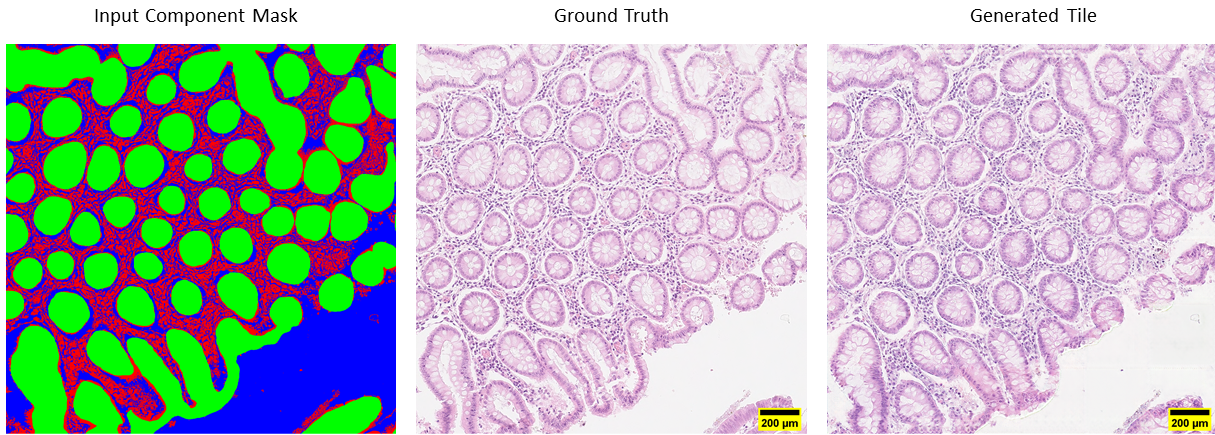}}
\caption{Generated tile of size $1436\times1436$ pixels from CRAG dataset}
\label{image_results_crag}
\end{figure*}

Fig. \ref{image_results_digestpath} shows generated tiles from the DigestPath test dataset after SAFRON is trained on the images extracted from the DigestPath train dataset. It shows generated tiles of sizes $5914\times4475$ and $6526\times5268$ along with a zoomed region from each tile. More high resolution generated tissue images can be found in the section \ref{generatedsamples} (in supplementary materials). For the high resolution version of these images, please visit the site\footnote{ https://warwick.ac.uk/TIALab/SAFRON}. For this purpose, the real tissue component mask of the image was used as input and this allows us to compare the generated image with the corresponding real image. We observe that, shapes, morphological characteristics and glandular appearances are preserved in the generated tile and the generated tile resembles the corresponding real image very closely. The quality of stitching operation is also shown in figure \ref{stitching}, where we can see seamless edge-crossing continuities. Though constructed tiles appear seamless and homogeneous, it can be noticed in the high resolution image that surface epithelium is not constructed with high fidelity. 
The visual results on a representative image from the CRAG test dataset are shown in Fig. \ref{image_results_crag}. Similar to the generated tile from the DigestPath dataset, we see that the generated tile appears seamless, preserving edge-crossing continuity and morphological characteristics of tissue including epithelial cells, glands and stroma. In addition, although epithelial cells and goblet cells can be clearly distinguished, some moderate deformities in glandular lumen portion are visible in the high resolution image.

\begin{figure*}
\centering
\begin{subfigure} {\columnwidth}
 \makebox[\textwidth]{\includegraphics[width=420pt,height=280pt]{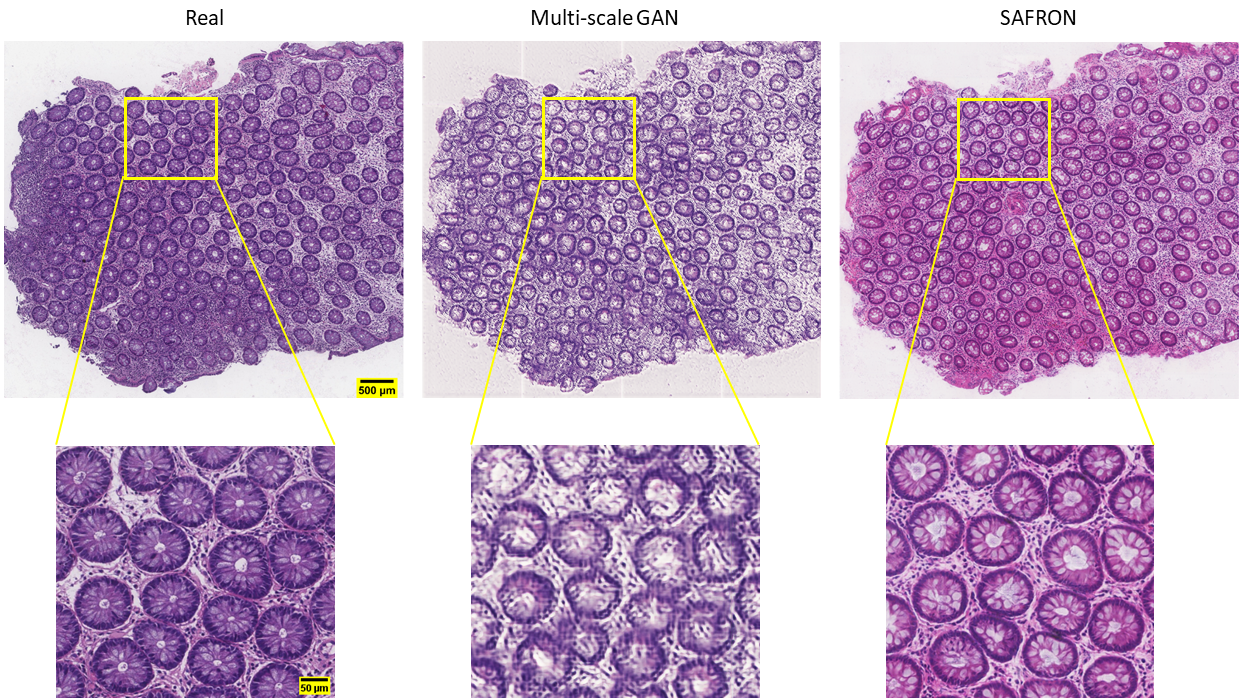}}
\caption{}
\end{subfigure}

\vspace{15pt}
\begin{subfigure} {\columnwidth}
 \makebox[\textwidth]{\includegraphics[width=400pt,height=140pt]{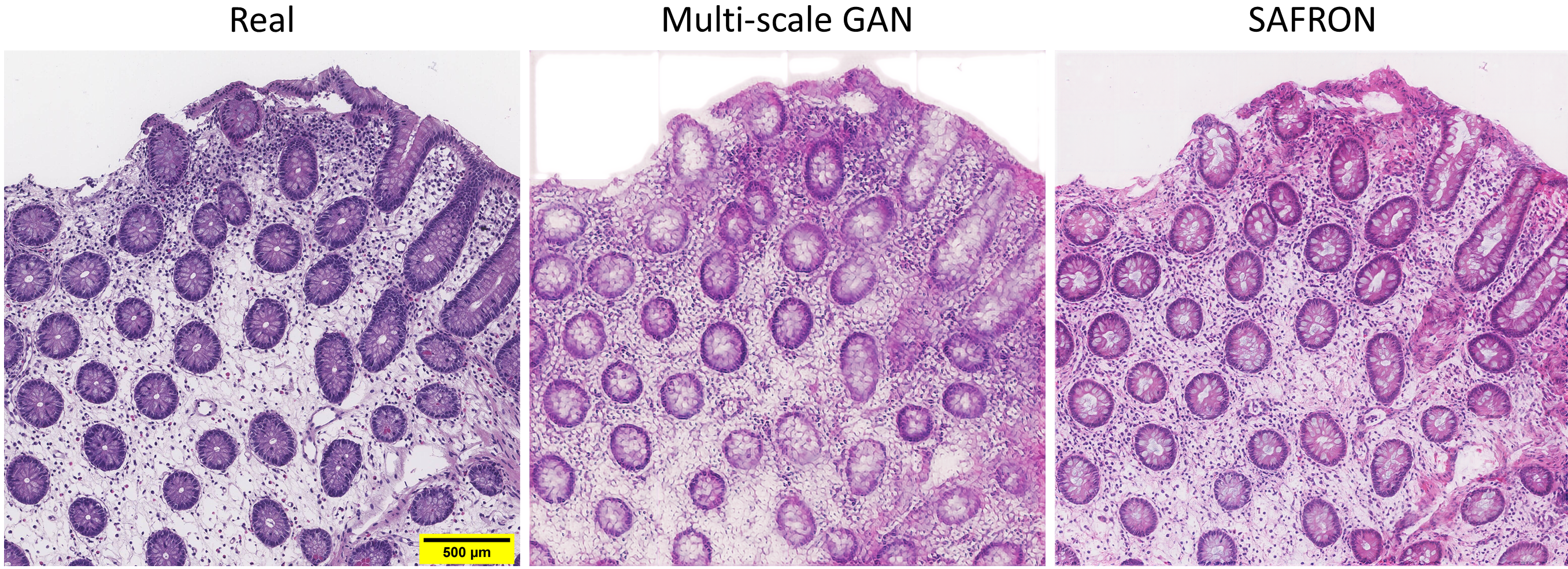}}
\caption{}
\end{subfigure}
\caption{Comparison of real image (on left) with generated images using Multi-scale GAN (on middle) and SAFRON (on right), after training them on the same data. The images on the top have resolution of $4096\times4096$ pixels, and those on the bottom have $2048\times2048$. The bottom row in (a) shows the zoomed areas from one of the parts of images.}
\label{multiscalegan_comparison}
\end{figure*}

Figure \ref{multiscalegan_comparison} shows the images generated via SAFRON and Multi-scale GAN frameworks from the DigestPath test set. It can be clearly noticed that tissue images generated by our SAFRON framework appear more closer to real images. The image generated by Multi-scale GAN appear loosing finer details like stromal region, nuclei, and goblet cells inside glands.

\begin{figure}[hbt!]
\centering
\includegraphics[width=250pt,height=90pt]{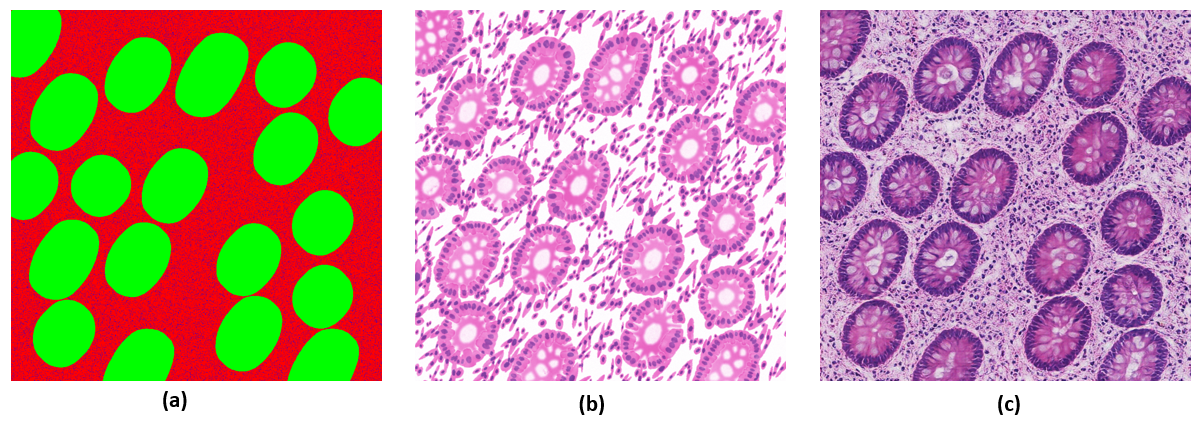}
\caption{From left to right: (a) tissue component mask, (b) synthetic tile generated by Kovacheva  et  al. \cite{kovacheva2016model}, (c) SAFRON method. All images have resolution of $1000\times1000$ pixels.}
\label{thecot_comparison}
\end{figure}

We have also compared the visual quality of generated images from SAFRON and the only other existing colorectal tissue image generation method THeCoT by Kovacheva et al. using the same tissue component mask for both methods. This comparison is shown in Fig. \ref{thecot_comparison}. The THeCoT parameters used for generating synthetic image are as follows: cellularity of stromal cells = 1, cellularity of epithelial cells = 1, grade of cancer = healthy. We can clearly see that tiles generated from SAFRON are significantly more realistic looking and have better perceptual quality.

\subsection{Quality Assessment with Frechet Inception Distance}
\label{fid_quality_section}

\begin{figure}[hbt!]
\centering
\includegraphics[width=180pt,height=140pt]{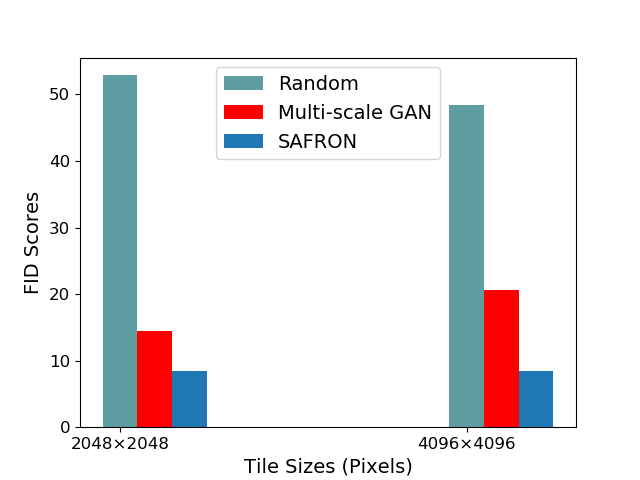}
\caption{Frechet Inception Distance comparison between SAFRON and Multi-scale GAN}
\label{fid_safron_msgan}
\end{figure}

To assess the proposed framework quantitatively, we compute the Frechet Inception Distance (FID) to evaluate the similarity between the sets of real and generated images. (\cite{heusel2017gans}) Lower FID corresponds to high perceptual and feature space similarity. For calculating FID, features from the last pooling layer of an Inception V3 network (\cite{szegedy2016rethinking}) trained on the ``ImageNet" database (\cite{deng2009imagenet}) were used.

For this experiment we extracted 128 tiles of size $4096\times4096$ and 226 tiles of size $2048\times2048$ from the DigestPath test set. Figure \ref{fid_safron_msgan} shows FID scores between a set of real and generated tiles computed by both SAFRON and baseline Multi-scale GAN frameworks. FID is heavily dependent upon the image size. So to get a sense of scale of FID, we also calculated the FID between real images and uniform random noise images of the same size. 

The figure \ref{fid_safron_msgan} reveals the lowest FID of 8.4 between real and SAFRON-generated tiles implies that the convolution feature maps computed from SAFRON-generated images are close to the ones obtained for real images. Consequently, it can be concluded that SAFRON-generated images are close to realistic images and can be used in computational pathology applications.

A unique quality of the SAFRON framework is that, it requires single training on relatively smaller images to generate images of large multiple resolutions. On the other hand, other frameworks including Multi-scale GAN require training data of the same resolution they target to generate, however acquisition of datasets having large number of high resolution annotated images is difficult. The generated tissue images of resolution $4096\times4096$ pixels using Multi-scale GAN resulted in degraded visual quality, and higher FID of 20.56, as compared to the slightly better quality, especially for the stromal regions, and lower FID of 14.42 obtained for the generated images of resolution $2048\times2048$ pixels. On the other hand, using only single time training on images of much smaller resolution $728\times728$ pixels, our SAFRON framework is able to generate high resolution images of arbitrary sizes, with higher fidelity, as well as smaller FID of 8.4 for $2048\times2048$ and 8.43 for $4096\times4096$ resolution images, respectively. All training images for both Multi-scale GAN and SAFRON were extracted from the same DigestPath train set.

\subsection{Quality Assessment through Gland Segmentation}

In this experiment, we use a U-Net \cite{ronneberger2015u} based gland segmentation model to quantify SAFRON image generation quality by comparing gland segmentation outputs for real and corresponding SAFRON-generated images. 

\begin{figure}[hbt!]
\centering
\includegraphics[width=250pt,height=60pt]{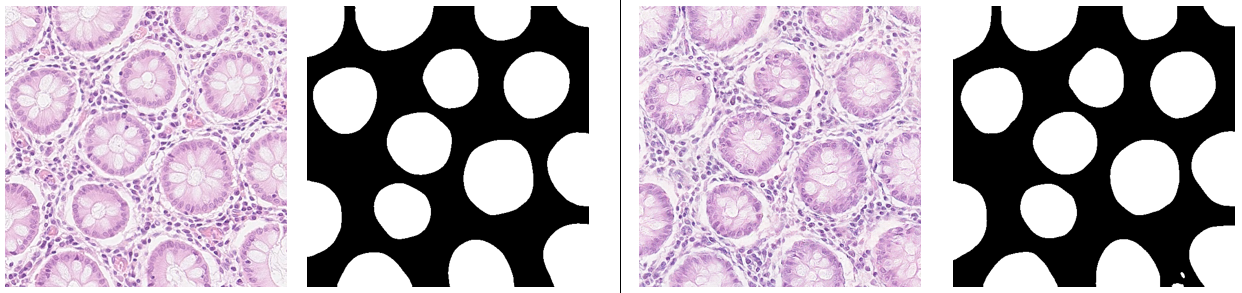}
\caption{Patches of original tiles (left) and generated tiles (right) along with their gland segmentation mask}
\label{unet_segmentation}
\end{figure}

For this purpose, we extracted 468 patches of size $512\times512$ from the CRAG training set to train a U-Net model for gland segmentation. Using the SAFRON framework trained on the CRAG training set, we computed synthetic counterparts of the images in the CRAG test set. We then extracted 471 patches of size $512\times512$ from real images as well as SAFRON-generated tiles from the CRAG test set, computed the binary gland segmentation mask on both sets using the trained U-Net model and calculated the Dice index score (\cite{zou2004statistical}) between the corresponding segmentation outputs as a quality metric. Sample results of segmentation on both real as well as synthetic patches can be seen in the Fig. \ref{unet_segmentation}.

We obtained an average Dice index score (\cite{zou2004statistical}) of 0.93 (with a standard deviation of 0.058) between the computed binary segmentation masks on patches of real and generated tiles. In comparison to an ideal Dice score of 1.0, the score of 0.93 obtained in this experiment shows that there is minimal difference between real and generated tiles for the purposes of gland segmentation. 

\subsection{Synthetic Images for Training \& Performance Evaluation of Machine Learning Models}

In this section we describe how SAFRON-generated images can be used for training and performance assessment of a U-Net based gland segmentation model~(\cite{ronneberger2015u}) with results at par with evaluation on real images. 

For this experiment, we extracted 72 $1436\times1436$ pixel tiles from the CRAG test set and divided them into two parts: 40 images for training (U-Net training tiles) and the remaining 32 images for testing (U-Net testing tiles). Using SAFRON trained on the CRAG train set, we generated synthetic counterparts of these 72 tiles. We then extracted two sets of 1930 $512\times512$ pixel patches - one from the 40 real U-Net training tiles and the other from SAFRON-generated counterparts of these images. These two sets are then used to train the same U-Net architecture to yield two different segmentation models - one trained on real images and the other one on SAFRON-generated synthetic images. Each U-Net model is trained with 80\% of the training images with 20\% used for internal validation with a batch size of 5 and a learning rate of $10^{-5}$.

\begin{table}[htb]
\centering
\begin{tabular}{|l|l|l|}
\hline
Testing\textbackslash{}Training & Real       & Synthetic   \\ \hline
Real                            & 0.91 (0.1) & 0.88 (0.14) \\ \hline
Synthetic                       & 0.90 (0.1)  & 0.97 (0.03) \\ \hline
\end{tabular}
\caption{Dice index scores obtained after training and testing U-Net on set of patches as shown under the respective column}
\label{traiability}
\end{table}

We then obtained two test sets of 1,639 $512\times512$ pixel patches each from the 32 U-Net test images and their SAFRON-generated counterparts. These test image patches are not used for training either SAFRON or the U-Net segmentation models and are used as unseen test sets for computing Dice scores of binarized segmentation masks outputs with respect to their corresponding ground truth data. 

The results of this experiment are shown in table \ref{traiability}. It shows that a U-Net gland segmentation model trained over real images yields an average Dice score of 0.91 using real test images and 0.90 over SAFRON-generated images. Similarly, a U-Net model trained over SAFRON-generated images and tested on real images gives an average Dice score of 0.88 in comparison to 0.97 when synthetic images are used for testing as well. These results clearly show that synthetic images can be used as a replacement of real images in training and performance evaluation of segmentation and possibly other methods in  computational pathology.
 
\subsection{Improving Data Efficiency with Synthetic Data}

In this section, we demonstrate how SAFRON-generated images can be used to overcome training data size limitations in training machine learning algorithms by improving their data efficiency. 

\begin{figure}[hbt!]
\centering
\includegraphics[width=250pt,height=170pt]{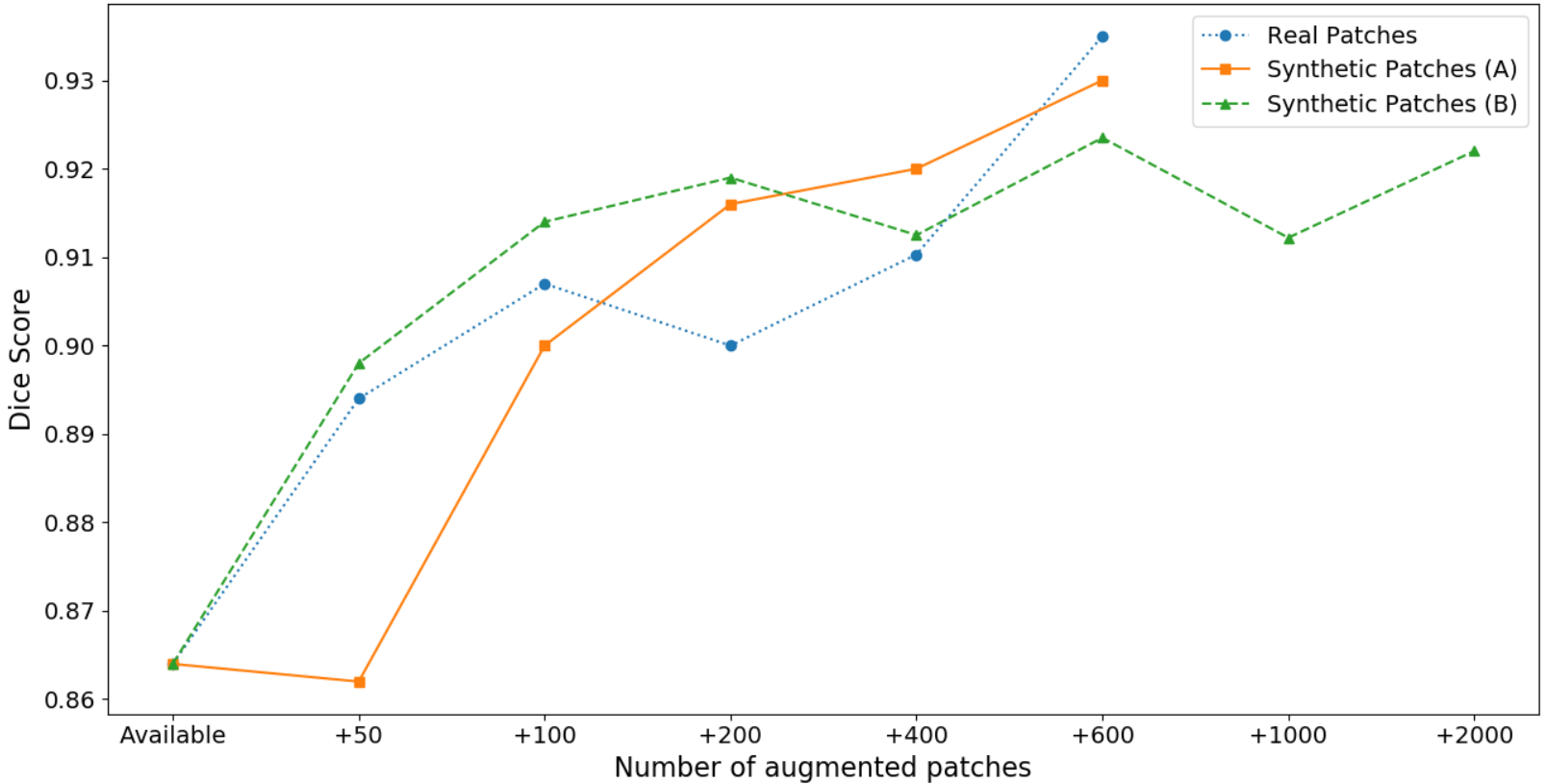}
\caption{Trend in Dice score after augmenting real and synthetic data. The augmented synthetic tissue patches are obtained from (A) component masks from Augmentation subset and (B) component masks generated using synthetic mask generation method, given in section 3.3}
\label{aug_data_exp}
\end{figure}

\begin{figure}[hbt!]
\centering
\includegraphics[width=170pt,height=95pt]{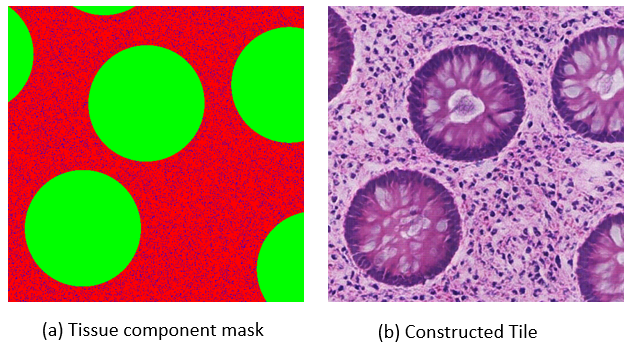}
\caption{Pair of tissue image along with its component mask constructed using synthetic mask generation method described in section 3.3. Both have resolution of $512\times512$ pixels.}
\label{random_tissue_gen}
\end{figure}

\begin{figure*}[hbt!] 
\centering
 \makebox[\textwidth]{\includegraphics[width=350pt,height=300pt]{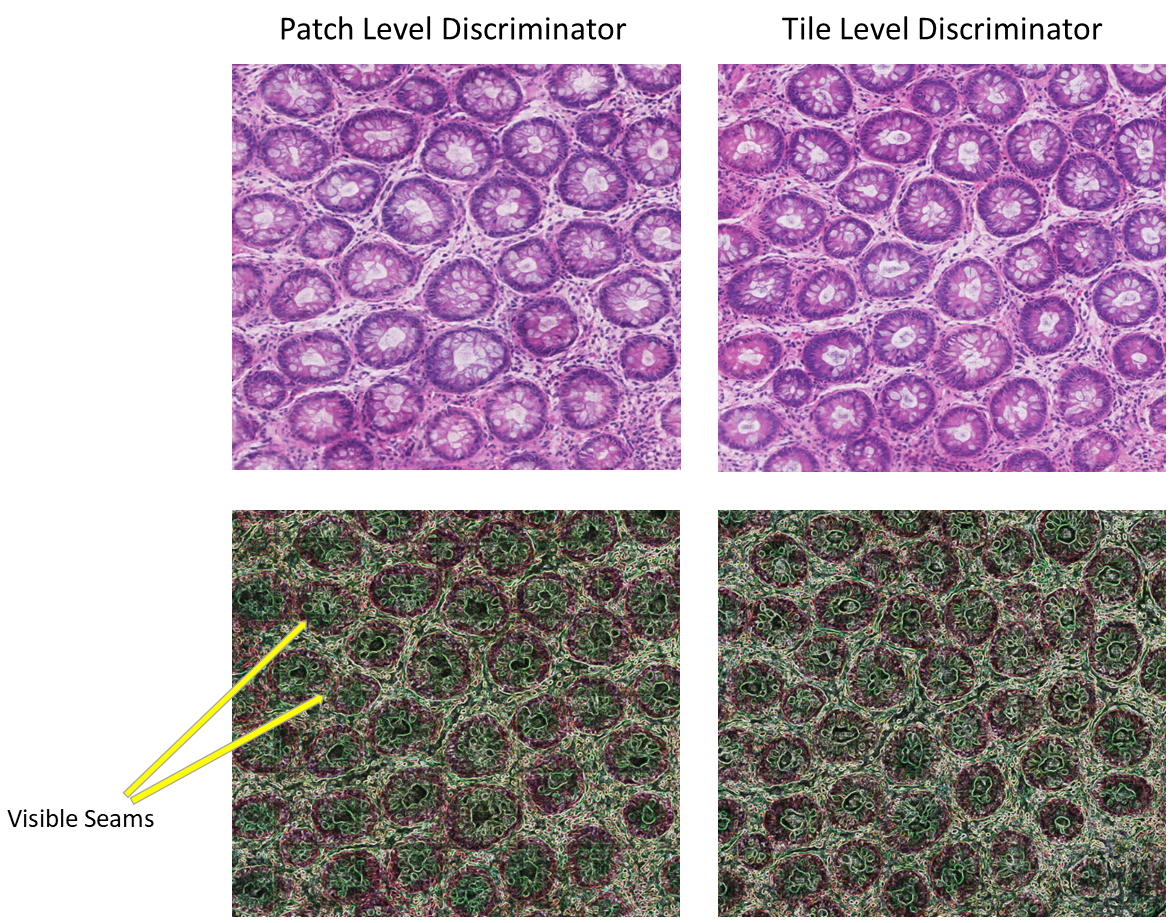}}
\caption{Ablation study showing importance of tile level generator. Top row shows the generated tissue tiles with the discriminator setting given on title. Bottom row shows the respective Sobel gradient images.}
\label{ablation}
\end{figure*}

In this experiment, we track the change in segmentation Dice score of a U-Net model over an unseen test set upon augmenting its training dataset with additional real and SAFRON-generated synthetic images. For this purpose, we first divided the 48 large images in the CRAG dataset into three equally sized non-overlapping subsets of 16 images called the Training subset, Test Subset, and Augmentation subset. The test subset is used only for testing the U-Net segmentation model and is not used for training either SAFRON or the U-Net gland segmentation model. As a baseline, we first trained the U-Net model on 144 patches of size $512\times512$ from the Training subset which yields a Dice score of 0.864 on patches from the test subset which is quite low due to the small amount of training data. We then train SAFRON on 640 square tiles of size $728\times728$ extracted from the same training subset and augment the U-Net training dataset with varying numbers of patches from two distinct types of SAFRON-generated synthetic images: one obtained using tissue component masks of images in the augmentation subset and the other based on the tissue component masks generated using synthetic mask generation method, as described in Section 3.3. The sample of generated tissue from synthetically generated component mask is shown in figure \ref{random_tissue_gen}.

The change in segmentation performance of U-Net models with different types and number of augmented patches is shown in figure \ref{aug_data_exp}. It shows that the addition of real images improves the Dice score from 0.864 to ~0.93 (with 500 additional patches). However, what is more interesting is the fact that the addition of synthetic images from SAFRON to the U-Net training dataset results in a similar improvement even when those images are obtained from randomly generated tissue component masks without using any additional real data either for training SAFRON or the segmentation model. 

This experiment clearly demonstrates the usefulness of the proposed scheme in cases where the amount of real annotated training data is small as is frequently the case in machine learning method development in computational pathology. 

\subsection{Ablation Studies}
\label{ablationstudy}

\subsubsection{Tile Level Discriminator}

We perform an ablation study to validate the seamless stitching of generated patches and global coherence across the final tile, by examining the importance of tile level discriminator used in the network. For this purpose, we train one variant of the SAFRON on same set of images extracted from DigestPath train set. In this variant, we use the PatchGAN discriminator used in (\cite{isola2017image}) that works on patch level and predicts the realism of the tissue patch generated by the generator. The final tile is constructed by spatially stitching of generated patches. Later we compute the Sobel gradients (\cite{sobel}) of the image samples extracted from images (from DigestPath test set) generated using these two networks (original network and its variant), to verify the seamless appearance of generated tiles. The gradients along with generated sample can be see in figure \ref{ablation}. We can clearly notice visible seams for the network with patch level discriminator (on the left of figure \ref{ablation}). The original network, with tile level discriminator shows great promise, which ascertain the model's architecture. This indicates the high applicability of tile level discriminator to generate seamless and globally coherent high resolution (large) images.

To verify it quantitatively, we compute the Frechet inception distance (FID) between the set of real and generated images taken from the stitching regions. The FID is calculated using features extracted from the last pooling layer of an Inception V3 network (\cite{szegedy2016rethinking}) trained on the ``ImageNet" database (\cite{deng2009imagenet}). For this experiment, we extracted 5000 image samples of small size of $76\times76$ symmetrically from the patches overlapping regions. The table \ref{ablation_fid} shows the results of FID calculated between set of real and generated samples, for both networks: original and its variant (described above), and also for randomly initialized images (to get a sense of scale of FID for this image size). The table \ref{ablation_fid} shows the considerably lesser value of the SAFRON with tile level discriminator which clearly conveys that, this design can be most beneficial to maintain the global coherence across tiles.

\begin{table}[htb!]
\centering
\begin{tabular}{|l|l|}
\hline
Model                     & FID             \\ \hline
Random                    & 1387.920        \\ \hline
Patch level discriminator & 227.538         \\ \hline
Tile level discriminator  & \textbf{142.84} \\ \hline
\end{tabular}
\caption{FID calculated from stitched regions}
\label{ablation_fid}
\end{table}

\begin{figure}[hbt!]
\centering
\includegraphics[width=260pt,height=100pt]{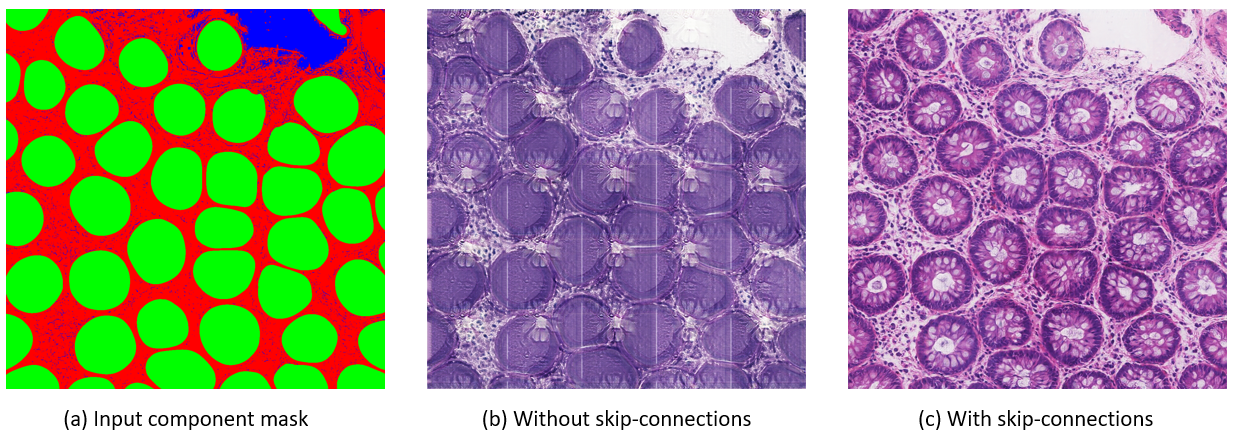}
\caption{Effect of Skip-Connections}
\label{skip-connections-compare}
\end{figure}

\begin{figure*}[bth!]
\centering
\includegraphics[width=\textwidth]{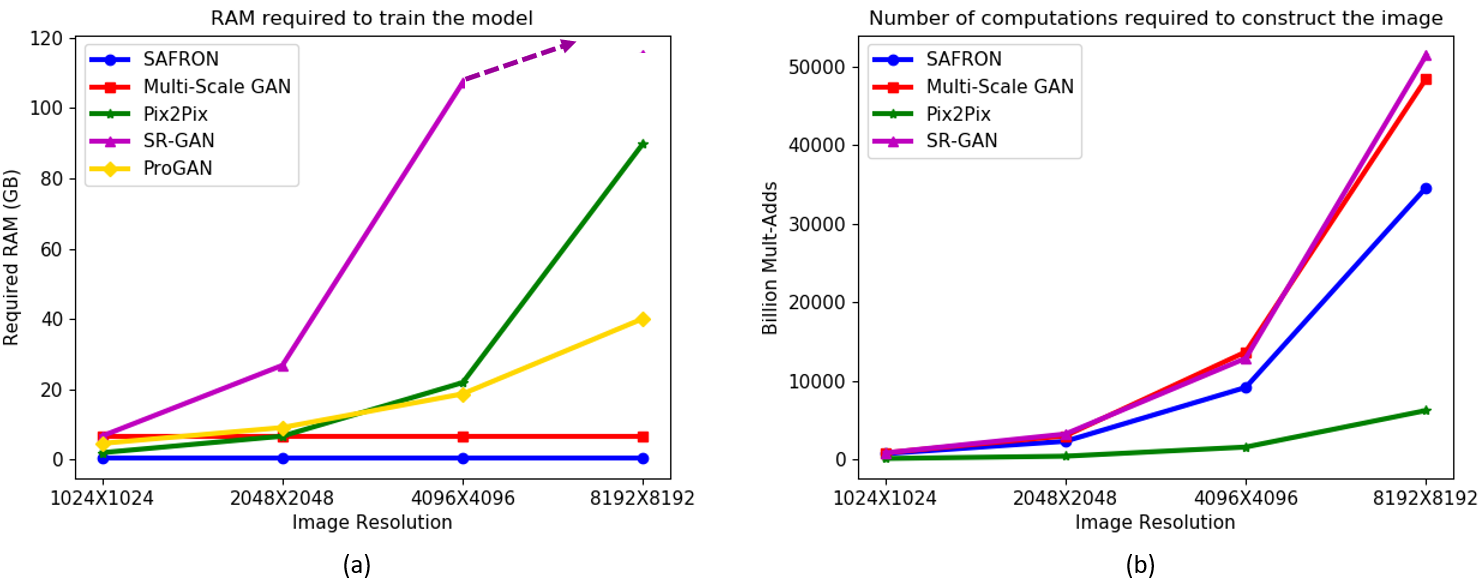}
\caption{The figure shows comparison of various models in terms of RAM demand for training the network (on the left) and number of computations (Mult-Adds) required to construct an image (on the right). The SAFRON framework requires constant memory of 0.35GB for training the model regardless of the image size. Dotted line (in (a)) shows extrapolated values of RAM demand for SR-GAN method for generation of images beyond the resolution of $4096\times4096$. We considered available RAM of 12GB for all experiments. Since Pro-GAN takes relatively very low computations to construct high resolution images, we didn't include it in (b). More details are given in section \ref{complexityanalysis}}
\label{ramdemand}
\end{figure*}

\subsubsection{Importance of Skip Connections}

In order to understand the effect of skip-connections used in the generator of the network, we evaluated the visual quality of generated images with and without skip-connections. 

Fig. \ref{skip-connections-compare} shows that the quality of generated images is significantly better when skip connections are used in comparison to when they are not. This shows that skip-connections in the SAFRON framework are crucial for the generation of high quality structures and glandular details.

\section{Memory Requirement and Runtime Analysis}
\label{complexityanalysis}

Due to computational and memory constraints, GANs are rarely applied for generation of high resolution images. In this experiment, we investigate the memory and computational requirements of the SAFRON framework. We present a comparative analysis between the SAFRON and other existing frameworks for generation of high resolution images.

We selected Multi-scale GAN (\cite{multiscalegan}), Pix2Pix (\cite{pix2pix}), SR-GAN (\cite{srgan}) and ProGAN (\cite{progan}) for comparison of memory and computational requirements. The Multi-scale GAN is adapted using the approach described in section \ref{results_exp_settings}. For Pix2pix, extra layers are added to the both generator and discriminator of the framework to generate high resolution images. The SR-GAN model up-scales the given image by factor of 4. We assumed a low resolution tissue image was already present in order to adapt this model. ProGAN is capable of generating high quality images and is considered for the generation of high resolution tissue images without using any mask. To increase the image to the desired resolution, additional layers are used. All models are implemented using same system settings used to implement the SAFRON as described in section \ref{results_exp_settings}.

\subsection{RAM Demand Computation}

RAM demand for training the framework is computed using the Keras summary approach (\cite{keras}). The calculated memory usage includes one forward and backward pass of the generator and discriminator each, and memory required to store all network parameters. The comparison of RAM demand to train models between SAFRON and other state-of-the-art networks for image generation is shown in figure \ref{ramdemand}(a). 

Figure \ref{ramdemand}(a) clearly shows that the RAM demand increases exponentially for Pix2Pix and SR-GAN with increase in the required image size. The methods Pix2pix, SRGAN and ProGAN can not be used to train models for generating images beyond resolution $2048 \times 2048$ considering the maximum RAM available of 12GB. These results indicate the infeasibility of those GAN based approaches towards generating high dimensional whole slide digital pathology images. The SR-GAN requires more than 100GB memory to train models on images with resolution $4096\times4096$, thus the extrapolated cubic curve is shown. In contrast, the proposed SAFRON requires constant 0.35 GB RAM for generating images of any resolution as the underlying generative process works at the patch-level. While Multi-scale GAN can generate large image patches but its memory requirements (6.73GB) is considerably higher in comparison to the proposed framework (0.35GB).

\subsection{Number of Computations}

In line with the computation complexity analysis done by (\cite{mobilenet}), we have also calculated the number of Mult-Adds (Multiply-Adds operations) operations required by various image generation frameworks Pix2Pix, PRO-GAN, Multi-Scale GAN, SR-GAN and compared them to SAFRON. Generally for neural network architectures, it is assumed that, 2 GFlops = 1 Mult-Adds where GFlops is giga floating point operations. In the SAFRON framework, construction of one patch of resolution $256\times256$ calculated to take 56.25 GFlops. To construct an image of resolution $M \times N$ and stride $S$, the number of patches is given by,

\begin{equation}
    N_{patches} = (M \times N)/S^2
\end{equation}

Thus, the total GFlops required to generate a final image is given by $N_{patches} \times 56.25$. Thus the computational complexity to construct a complete image is dependent directly on the number of patches. 

For comparative analysis, we calculated the required Mult-Adds to construct a single image of various resolutions. As discussed earlier, only SAFRON and Multi-scale GAN have memory requirements under 12GB making other frameworks unusable under practical settings for generating large image patches. Therefore we initialized the remaining frameworks with dummy random weights to calculate the required number of computations. ProGAN needs around 70 billion Mult-Adds to construct and image of resolution $8192\times8192$. This is because using up-scaling and down-scaling operations in place of transpose and regular convolutions respectively, which needs considerably lesser number of computations. The plots of number of computations, which excludes ProGAN can be seen in figure \ref{ramdemand}(b). As the number of patches increases with the size, we can see an increasing curve of required billion Mult-Adds for our framework, but much lesser than those required by Multi-Scale GAN framework. 

\begin{figure}[hbt!]
\centering
\includegraphics[width=250pt,height=200pt]{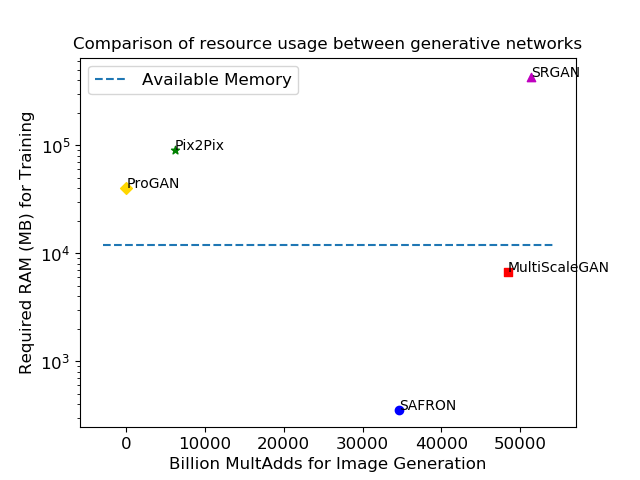}
\caption{Resource usage chart required to construct and image of size $8192\times8192$. Dotted line shows the available memory of 12GB. Our framework and Multi-Scale GAN appear feasible for training to generate high resolution images.}
\label{resource}
\end{figure}

The figures \ref{resource} shows the entire picture of resource usage to generate an image of resolution $8192\times8192$, in terms of number of computations and RAM requirement by all of the considered frameworks including the proposed SAFRON method.

\begin{figure}[hbt!]
\centering
\includegraphics[width=250pt,height=200pt]{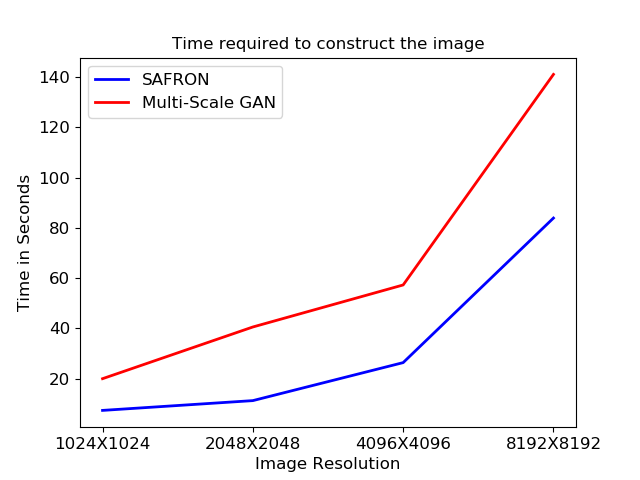}
\caption{Time required to construct high resolution images using the SAFRON and Multi-scale GAN}
\label{time}
\end{figure}

The blue dotted line is the line of feasibility under which the models are implementable considering the available memory of 12GB. Though Pix2Pix and ProGAN appears to take less computations to construct the image, but they still can not be implemented in the given system due to high memory requirement. Apart from Multi-Scale GAN and our framework, the construction of whole slide images which are multi-gigapixel in nature, are not possible with other existing frameworks. 

We calculated the required time to construct high resolution images by ours and Multi-Scale GAN. The plot \ref{time} shows the required time to construct an image of increasing resolutions.  Empirically we found that, our framework took less than 2 minutes to construct an image of resolution $8000 \times 8000$. The time requirement is not only dependent on number of computations, but on the model architecture as well. Multi-scale GANs operates several GANs in series, with each of the network is responsible for increasing resolution of images generated in previous layers. Our framework is able to perform patch-wise processing in parallel before final stitching operation to get entire high resolution image. Thus, our framework gets an edge over Multi-scale GANs in terms of the time requirement.

\section{Discussion}

The computationally and memory efficient SAFRON framework can be thought as a crucial step towards construction of complete whole slide images. Whole slide images are generally multi-gigapixel in nature and have great mixture of various tissue components with large number of inter-dependencies among them. The tissue component masks for such images can be generated or acquired using segmentation techniques depending on digital pathology algorithms. The framework can potentially be useful in generating an exhaustive annotated data for training and evaluation of classification/segmentation task like the gland segmentation in the domain of computational histopathology.

Moreover, the discriminator output can be utilized to compute importance weights as given in (\cite{robustlabelsynthesize}). The weighted training can potentially increase the performance for segmentation/classification algorithms. However, it is difficult to achieve importance weights for high resolution images from the discriminator neural network, which consumes an input image of limited size ($728\times728$ in our case) while training. One can aggregate the importance weights computed from patches of high resolution images to compute the final weight.

As the generator used in SAFRON framework constructs tissue image patches from the local area of component mask, the SAFRON framework design is appropriate for generating digital pathology images. We keep the component mask patch size greater than the target tissue patch size in order to encourage the context aware generation. For tissue images, this local context appears to be sufficient to generate tissue image patches, which later get stitched to generate the final tile.

\section{Conclusions \& Future Directions}

We presented a novel framework, SAFRON, for generation of large histology synthetic image tiles while training on relatively smaller image patches. The constructed tiles from the patch-based framework do not exhibit any boundary artifacts or other deformities between adjacent patches. We showed that the synthetic tissue image tiles generated by SAFRON preserve morphological characteristics in the tissue regions, as demonstrated by the high Dice index for gland segmentation. We illustrated the computational efficiency of the SAFRON framework in terms of RAM demand for training the model and number of computations to construct an image, compared to existing approaches for generation of high resolution images. We assessed the quality of the generated tissue images through the FID metric, using the convolutional features of an Inception-V3 network. The images generated using SAFRON showed the consistent FID of around 8.4 and outperformed those generated by the baseline Multi-scale GAN with a significant margin. We demonstrated in this work that the synthetic image tiles constructed from our framework accompanied with a definitive ground-truth generated by a parametric model can be used for pre-training and evaluation of deep learning algorithms in circumstances where labeled data is scarce.

SAFRON may be used to extend already existing segmentation datasets for histology image analysis, enabling researchers to improve the performance of automated segmentation approaches for computational pathology. This framework can be generalized for producing a large number of tiles for different types of carcinomas and tissue types. An open future direction for research would be to extend the SAFRON framework to generate complete whole-slide images from known parameters.

\section*{Acknowledgements}
Fayyaz Minhas, Simon Graham and Nasir Rajpoot are part of the PathLAKE digital pathology consortium, which is funded from the Data to Early Diagnosis and Precision Medicine strand of the governments Industrial Strategy Challenge Fund, managed and delivered by UK Research and Innovation (UKRI). Nasir Rajpoot was also supported by the UK Medical Research Council (grant award MR/P015476/1) and the Alan Turing Institute.

\vspace{60pt}

\newpage

\bibliographystyle{IEEEtran}
\bibliography{ref}

\begin{thebibliography}{10}
\providecommand{\url}[1]{#1}
\csname url@samestyle\endcsname
\providecommand{\newblock}{\relax}
\providecommand{\bibinfo}[2]{#2}
\providecommand{\BIBentrySTDinterwordspacing}{\spaceskip=0pt\relax}
\providecommand{\BIBentryALTinterwordstretchfactor}{4}
\providecommand{\BIBentryALTinterwordspacing}{\spaceskip=\fontdimen2\font plus
\BIBentryALTinterwordstretchfactor\fontdimen3\font minus
  \fontdimen4\font\relax}
\providecommand{\BIBforeignlanguage}[2]{{%
\expandafter\ifx\csname l@#1\endcsname\relax
\typeout{** WARNING: IEEEtran.bst: No hyphenation pattern has been}%
\typeout{** loaded for the language `#1'. Using the pattern for}%
\typeout{** the default language instead.}%
\else
\language=\csname l@#1\endcsname
\fi
#2}}
\providecommand{\BIBdecl}{\relax}
\BIBdecl

\bibitem{tumorsegment1}
\BIBentryALTinterwordspacing
Y.~Xu, J.-Y. Zhu, E.~I.-C. Chang, M.~Lai, and Z.~Tu, ``Weakly supervised
  histopathology cancer image segmentation and classification,'' \emph{Medical
  Image Analysis}, vol.~18, no.~3, pp. 591 -- 604, 2014. [Online]. Available:
  \url{http://www.sciencedirect.com/science/article/pii/S1361841514000188}
\BIBentrySTDinterwordspacing

\bibitem{bejnordi2017diagnostic}
B.~E. Bejnordi, M.~Veta, P.~J. Van~Diest, B.~Van~Ginneken, N.~Karssemeijer,
  G.~Litjens, J.~A. Van Der~Laak, M.~Hermsen, Q.~F. Manson, M.~Balkenhol
  \emph{et~al.}, ``Diagnostic assessment of deep learning algorithms for
  detection of lymph node metastases in women with breast cancer,''
  \emph{Jama}, vol. 318, no.~22, pp. 2199--2210, 2017.

\bibitem{qaiser2019fast}
T.~Qaiser, Y.-W. Tsang, D.~Taniyama, N.~Sakamoto, K.~Nakane, D.~Epstein, and
  N.~Rajpoot, ``Fast and accurate tumor segmentation of histology images using
  persistent homology and deep convolutional features,'' \emph{Medical image
  analysis}, vol.~55, pp. 1--14, 2019.

\bibitem{graham2019mild}
S.~Graham, H.~Chen, J.~Gamper, Q.~Dou, P.-A. Heng, D.~Snead, Y.~W. Tsang, and
  N.~Rajpoot, ``Mild-net: minimal information loss dilated network for gland
  instance segmentation in colon histology images,'' \emph{Medical image
  analysis}, vol.~52, pp. 199--211, 2019.

\bibitem{chen2017dcan}
H.~Chen, X.~Qi, L.~Yu, Q.~Dou, J.~Qin, and P.-A. Heng, ``Dcan: Deep
  contour-aware networks for object instance segmentation from histology
  images,'' \emph{Medical image analysis}, vol.~36, pp. 135--146, 2017.

\bibitem{cgrading1}
P.~Gupta, S.-F. Chiang, P.~Sahoo, S.~Mohapatra, J.-F. You, D.~Onthoni, H.-Y.
  Hung, J.-M. Chiang, Y.~Huang, and W.-S. Tsai, ``Prediction of colon cancer
  stages and survival period with machine learning approach,'' \emph{Cancers},
  vol.~11, 12 2019.

\bibitem{shaban2020context}
M.~Shaban, R.~Awan, M.~M. Fraz, A.~Azam, Y.-W. Tsang, D.~Snead, and N.~M.
  Rajpoot, ``Context-aware convolutional neural network for grading of
  colorectal cancer histology images,'' \emph{IEEE Transactions on Medical
  Imaging}, 2020.

\bibitem{zhou2019cgc}
Y.~Zhou, S.~Graham, N.~Alemi~Koohbanani, M.~Shaban, P.-A. Heng, and N.~Rajpoot,
  ``Cgc-net: Cell graph convolutional network for grading of colorectal cancer
  histology images,'' in \emph{Proceedings of the IEEE International Conference
  on Computer Vision Workshops}, 2019, pp. 0--0.

\bibitem{ncd1}
\BIBentryALTinterwordspacing
S.~Tripathi and S.~K. Singh, ``Cell nuclei classification in histopathological
  images using hybrid o<sub>l</sub>convnet,'' \emph{ACM Trans. Multimedia
  Comput. Commun. Appl.}, vol.~16, no.~1s, Mar. 2020. [Online]. Available:
  \url{https://doi.org/10.1145/3345318}
\BIBentrySTDinterwordspacing

\bibitem{graham2018sams}
S.~Graham and N.~M. Rajpoot, ``Sams-net: Stain-aware multi-scale network for
  instance-based nuclei segmentation in histology images,'' in \emph{2018 IEEE
  15th International Symposium on Biomedical Imaging (ISBI 2018)}.\hskip 1em
  plus 0.5em minus 0.4em\relax IEEE, 2018, pp. 590--594.

\bibitem{graham2019hover}
S.~Graham, Q.~D. Vu, S.~E.~A. Raza, A.~Azam, Y.~W. Tsang, J.~T. Kwak, and
  N.~Rajpoot, ``Hover-net: Simultaneous segmentation and classification of
  nuclei in multi-tissue histology images,'' \emph{Medical Image Analysis},
  vol.~58, p. 101563, 2019.

\bibitem{sirinukunwattana2016locality}
K.~Sirinukunwattana, S.~E.~A. Raza, Y.-W. Tsang, D.~R. Snead, I.~A. Cree, and
  N.~M. Rajpoot, ``Locality sensitive deep learning for detection and
  classification of nuclei in routine colon cancer histology images,''
  \emph{IEEE transactions on medical imaging}, vol.~35, no.~5, pp. 1196--1206,
  2016.

\bibitem{senaras2018optimized}
C.~Senaras, M.~K.~K. Niazi, B.~Sahiner, M.~P. Pennell, G.~Tozbikian,
  G.~Lozanski, and M.~N. Gurcan, ``Optimized generation of high-resolution
  phantom images using cgan: Application to quantification of ki67 breast
  cancer images,'' \emph{PloS one}, vol.~13, no.~5, p. e0196846, 2018.

\bibitem{senaras2018creating}
C.~Senaras, B.~Sahiner, G.~Tozbikian, G.~Lozanski, and M.~N. Gurcan, ``Creating
  synthetic digital slides using conditional generative adversarial networks:
  application to ki67 staining,'' in \emph{Medical Imaging 2018: Digital
  Pathology}, vol. 10581.\hskip 1em plus 0.5em minus 0.4em\relax International
  Society for Optics and Photonics, 2018, p. 1058103.

\bibitem{quiros2019pathology}
A.~C. Quiros, R.~Murray-Smith, and K.~Yuan, ``Pathology gan: learning deep
  representations of cancer tissue,'' \emph{arXiv preprint arXiv:1907.02644},
  2019.

\bibitem{bejnordi2017context}
B.~E. Bejnordi, G.~Zuidhof, M.~Balkenhol, M.~Hermsen, P.~Bult, B.~van Ginneken,
  N.~Karssemeijer, G.~Litjens, and J.~van~der Laak, ``Context-aware stacked
  convolutional neural networks for classification of breast carcinomas in
  whole-slide histopathology images,'' \emph{Journal of Medical Imaging},
  vol.~4, no.~4, p. 044504, 2017.

\bibitem{jacobkathermsi}
J.~Krause, H.~Grabsch, M.~Kloor, M.~Jendrusch, A.~Echle, R.~Buelow, P.~Boor,
  T.~Luedde, T.~Brinker, C.~Trautwein, A.~Pearson, P.~Quirke, J.~Jenniskens,
  K.~Offermans, P.~Brandt, and J.~Kather, ``Deep learning detects genetic
  alterations in cancer histology generated by adversarial networks,''
  \emph{The Journal of pathology}, 02 2021.

\bibitem{sashimi}
S.~Deshpande, F.~Minhas, and N.~Rajpoot, ``Train small, generate big: Synthesis
  of colorectal cancer histology images,'' in \emph{Simulation and Synthesis in
  Medical Imaging}, N.~Burgos, D.~Svoboda, J.~M. Wolterink, and C.~Zhao,
  Eds.\hskip 1em plus 0.5em minus 0.4em\relax Cham: Springer International
  Publishing, 2020, pp. 164--173.

\bibitem{digestpath}
J.~Li, S.~Yang, X.~Huang, Q.~Da, X.~Yang, Z.~Hu, Q.~Duan, C.~Wang, and H.~Li,
  \emph{Signet Ring Cell Detection with a Semi-supervised Learning Framework},
  05 2019, pp. 842--854.

\bibitem{simucell}
S.~Rajaram, B.~Pavie, N.~Hac, S.~Altschuler, and L.~Wu, ``Simucell: A flexible
  framework for creating synthetic microscopy images,'' \emph{Nature methods},
  vol.~9, pp. 634--5, 06 2012.

\bibitem{zhao2007automated}
T.~Zhao and R.~F. Murphy, ``Automated learning of generative models for
  subcellular location: building blocks for systems biology,'' \emph{Cytometry
  part A}, vol.~71, no.~12, pp. 978--990, 2007.

\bibitem{kovacheva2016model}
V.~N. Kovacheva, D.~Snead, and N.~M. Rajpoot, ``A model of the spatial tumour
  heterogeneity in colorectal adenocarcinoma tissue,'' \emph{BMC
  bioinformatics}, vol.~17, no.~1, p. 255, 2016.

\bibitem{goodfellow2014generative}
I.~Goodfellow, J.~Pouget-Abadie, M.~Mirza, B.~Xu, D.~Warde-Farley, S.~Ozair,
  A.~Courville, and Y.~Bengio, ``Generative adversarial nets,'' in
  \emph{Advances in neural information processing systems}, 2014, pp.
  2672--2680.

\bibitem{medicalgan}
Q.~Zhang, H.~Wang, H.~Lu, D.~Won, and S.~W. Yoon, ``Medical image synthesis
  with generative adversarial networks for tissue recognition,'' 06 2018.

\bibitem{robustlabelsynthesize}
L.~Hou, A.~Agarwal, D.~Samaras, T.~Kurc, R.~Gupta, and J.~Saltz, ``Robust
  histopathology image analysis: To label or to synthesize?'' 06 2019, pp.
  8525--8534.

\bibitem{mirza2014conditional}
M.~Mirza and S.~Osindero, ``Conditional generative adversarial nets,''
  \emph{arXiv preprint arXiv:1411.1784}, 2014.

\bibitem{progan}
T.~Karras, T.~Aila, S.~Laine, and J.~Lehtinen, ``Progressive growing of gans
  for improved quality, stability, and variation,'' 10 2017.

\bibitem{srgan}
C.~{Ledig}, L.~{Theis}, F.~{Huszár}, J.~{Caballero}, A.~{Cunningham},
  A.~{Acosta}, A.~{Aitken}, A.~{Tejani}, J.~{Totz}, Z.~{Wang}, and W.~{Shi},
  ``Photo-realistic single image super-resolution using a generative
  adversarial network,'' in \emph{2017 IEEE Conference on Computer Vision and
  Pattern Recognition (CVPR)}, 2017, pp. 105--114.

\bibitem{multiscalegan}
H.~Uzunova, J.~Ehrhardt, F.~Jacob, A.~Frydrychowicz, and H.~Handels,
  ``Multi-scale gans for memory-efficient generation of high resolution medical
  images,'' in \emph{Medical Image Computing and Computer Assisted Intervention
  -- MICCAI 2019}, D.~Shen, T.~Liu, T.~M. Peters, L.~H. Staib, C.~Essert,
  S.~Zhou, P.-T. Yap, and A.~Khan, Eds.\hskip 1em plus 0.5em minus 0.4em\relax
  Cham: Springer International Publishing, 2019, pp. 112--120.

\bibitem{kingma2019introduction}
D.~P. Kingma and M.~Welling, ``An introduction to variational autoencoders,''
  \emph{arXiv preprint arXiv:1906.02691}, 2019.

\bibitem{awan2017glandular}
R.~Awan, K.~Sirinukunwattana, D.~Epstein, S.~Jefferyes, U.~Qidwai, Z.~Aftab,
  I.~Mujeeb, D.~Snead, and N.~Rajpoot, ``Glandular morphometrics for objective
  grading of colorectal adenocarcinoma histology images,'' \emph{Scientific
  reports}, vol.~7, no.~1, pp. 1--12, 2017.

\bibitem{graham2020dense}
S.~Graham, D.~Epstein, and N.~Rajpoot, ``Dense steerable filter cnns for
  exploiting rotational symmetry in histology images,'' \emph{arXiv preprint
  arXiv:2004.03037}, 2020.

\bibitem{ding2020multi}
H.~Ding, Z.~Pan, Q.~Cen, Y.~Li, and S.~Chen, ``Multi-scale fully convolutional
  network for gland segmentation using three-class classification,''
  \emph{Neurocomputing}, vol. 380, pp. 150--161, 2020.

\bibitem{glas1}
K.~Sirinukunwattana, J.~P. Pluim, H.~Chen, X.~Qi, P.-A. Heng, Y.~B. Guo, L.~Y.
  Wang, B.~J. Matuszewski, E.~Bruni, U.~Sanchez, A.~Böhm, O.~Ronneberger,
  B.~B. Cheikh, D.~Racoceanu, P.~Kainz, M.~Pfeiffer, M.~Urschler, D.~R. Snead,
  and N.~M. Rajpoot, ``Gland segmentation in colon histology images: The glas
  challenge contest,'' \emph{Medical Image Analysis}, vol.~35, pp. 489--502,
  jan 2017.

\bibitem{glas2}
K.~{Sirinukunwattana}, D.~R.~J. {Snead}, and N.~M. {Rajpoot}, ``A stochastic
  polygons model for glandular structures in colon histology images,''
  \emph{IEEE Transactions on Medical Imaging}, vol.~34, no.~11, pp. 2366--2378,
  2015.

\bibitem{ronneberger2015u}
O.~Ronneberger, P.~Fischer, and T.~Brox, ``U-net: Convolutional networks for
  biomedical image segmentation,'' vol. 9351, 10 2015, pp. 234--241.

\bibitem{isola2017image}
P.~{Isola}, J.~{Zhu}, T.~{Zhou}, and A.~A. {Efros}, ``Image-to-image
  translation with conditional adversarial networks,'' in \emph{2017 IEEE
  Conference on Computer Vision and Pattern Recognition (CVPR)}, 2017, pp.
  5967--5976.

\bibitem{pix2pix}
P.~Isola, J.-Y. Zhu, T.~Zhou, and A.~A. Efros, ``Image-to-image translation
  with conditional adversarial networks,'' in \emph{Proceedings of the IEEE
  conference on computer vision and pattern recognition}, 2017, pp. 1125--1134.

\bibitem{tensorflow}
\BIBentryALTinterwordspacing
M.~Abadi, A.~Agarwal, P.~Barham, E.~Brevdo, Z.~Chen, C.~Citro, G.~S. Corrado,
  A.~Davis, J.~Dean, M.~Devin, S.~Ghemawat, I.~Goodfellow, A.~Harp, G.~Irving,
  M.~Isard, Y.~Jia, R.~Jozefowicz, L.~Kaiser, M.~Kudlur, J.~Levenberg,
  D.~Man\'{e}, R.~Monga, S.~Moore, D.~Murray, C.~Olah, M.~Schuster, J.~Shlens,
  B.~Steiner, I.~Sutskever, K.~Talwar, P.~Tucker, V.~Vanhoucke, V.~Vasudevan,
  F.~Vi\'{e}gas, O.~Vinyals, P.~Warden, M.~Wattenberg, M.~Wicke, Y.~Yu, and
  X.~Zheng, ``{TensorFlow}: Large-scale machine learning on heterogeneous
  systems,'' 2015, software available from tensorflow.org. [Online]. Available:
  \url{https://www.tensorflow.org/}
\BIBentrySTDinterwordspacing

\bibitem{keras}
F.~Chollet \emph{et~al.}, ``Keras,'' \url{https://github.com/fchollet/keras},
  2015.

\bibitem{heusel2017gans}
M.~Heusel, H.~Ramsauer, T.~Unterthiner, B.~Nessler, and S.~Hochreiter, ``Gans
  trained by a two time-scale update rule converge to a local nash
  equilibrium,'' in \emph{Advances in neural information processing systems},
  2017, pp. 6626--6637.

\bibitem{szegedy2016rethinking}
C.~Szegedy, V.~Vanhoucke, S.~Ioffe, J.~Shlens, and Z.~Wojna, ``Rethinking the
  inception architecture for computer vision,'' in \emph{Proceedings of the
  IEEE conference on computer vision and pattern recognition}, 2016, pp.
  2818--2826.

\bibitem{deng2009imagenet}
J.~Deng, W.~Dong, R.~Socher, L.-J. Li, K.~Li, and L.~Fei-Fei, ``Imagenet: A
  large-scale hierarchical image database,'' in \emph{2009 IEEE conference on
  computer vision and pattern recognition}.\hskip 1em plus 0.5em minus
  0.4em\relax Ieee, 2009, pp. 248--255.

\bibitem{zou2004statistical}
K.~H. Zou, S.~K. Warfield, A.~Bharatha, C.~M. Tempany, M.~R. Kaus, S.~J. Haker,
  W.~M. Wells~III, F.~A. Jolesz, and R.~Kikinis, ``Statistical validation of
  image segmentation quality based on a spatial overlap index1: scientific
  reports,'' \emph{Academic radiology}, vol.~11, no.~2, pp. 178--189, 2004.

\bibitem{sobel}
I.~Sobel, ``An isotropic 3x3 image gradient operator,'' \emph{Presentation at
  Stanford A.I. Project 1968}, 02 2014.

\bibitem{mobilenet}
A.~Howard, M.~Zhu, B.~Chen, D.~Kalenichenko, W.~Wang, T.~Weyand, M.~Andreetto,
  and H.~Adam, ``Mobilenets: Efficient convolutional neural networks for mobile
  vision applications,'' 04 2017.

\end{thebibliography}

\newpage

\section{Supplementary Material}

\subsection{Encoder Decoder Architecture}

\begin{figure}[!hbpt]
\centering
\includegraphics[width=250pt,height=120pt]{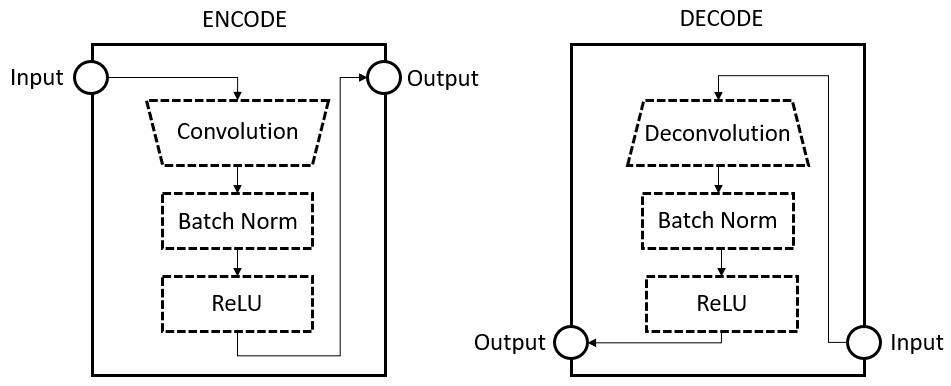}
\caption{Encoder Decoder Architecture}
\label{encodedecode}
\end{figure}

\subsection{Generated Samples of Various Resolutions}
\label{generatedsamples}

\begin{figure}[!hbpt]
\centering
\includegraphics[width=350pt,height=350pt]{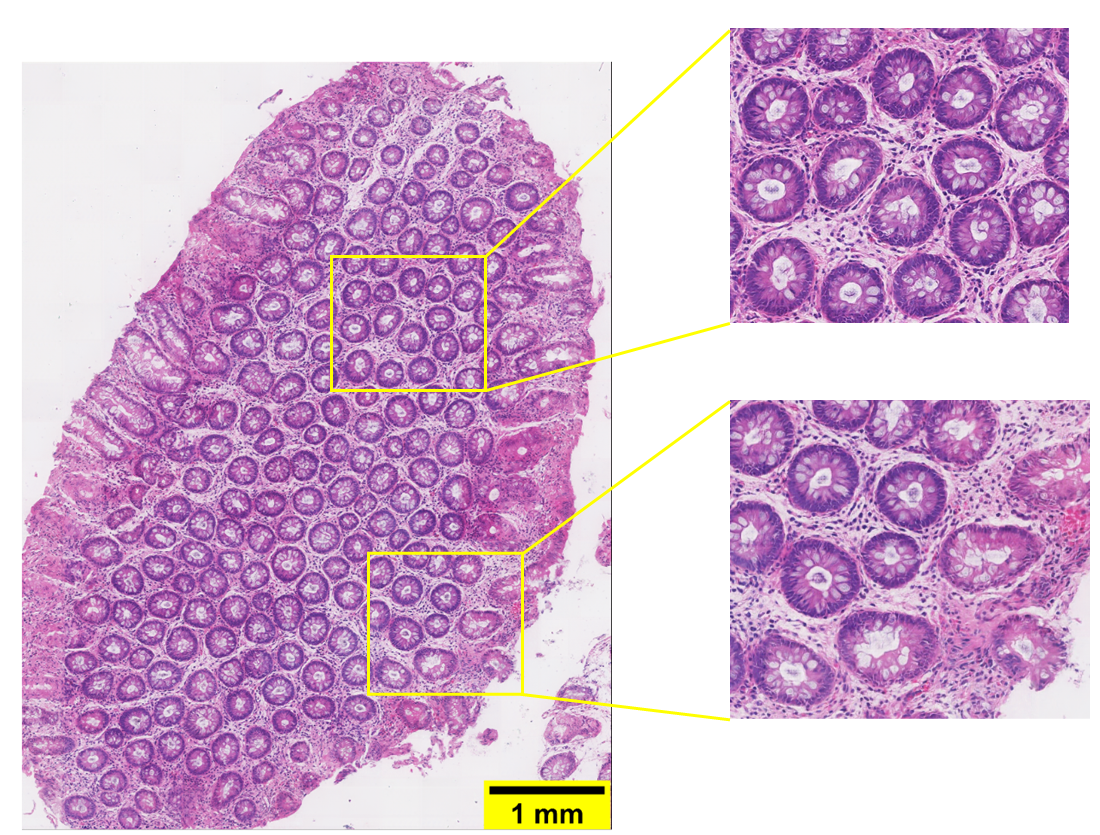}
\caption{Sample of generated image of size $3546\times4610$ Pixels}
\end{figure}

\begin{figure*}[!hbpt]
\centering
\includegraphics[width=450pt,height=300pt]{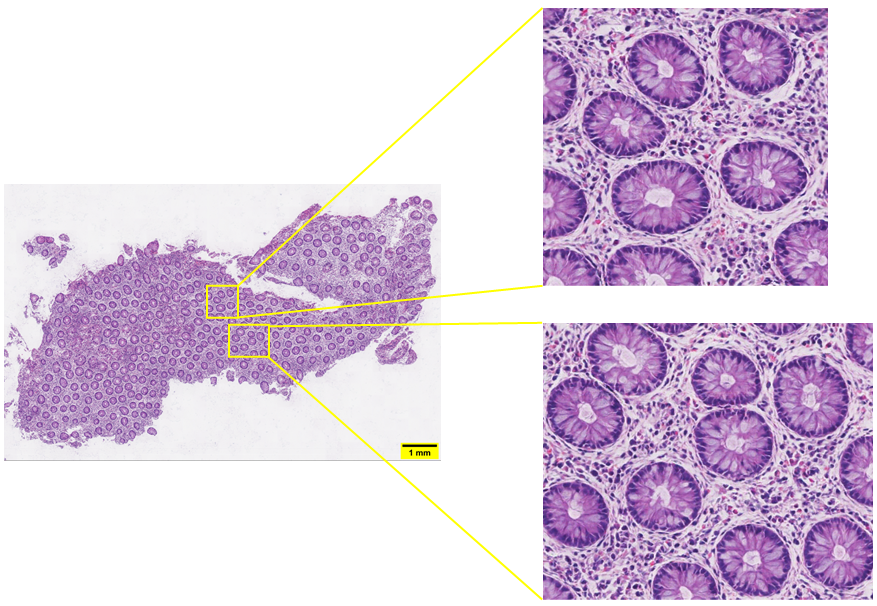}
\caption{Sample of generated image of size $8743\times5611$ Pixels}
\end{figure*}

\begin{figure*}[!hbpt]
\centering
\includegraphics[width=400pt,height=320pt]{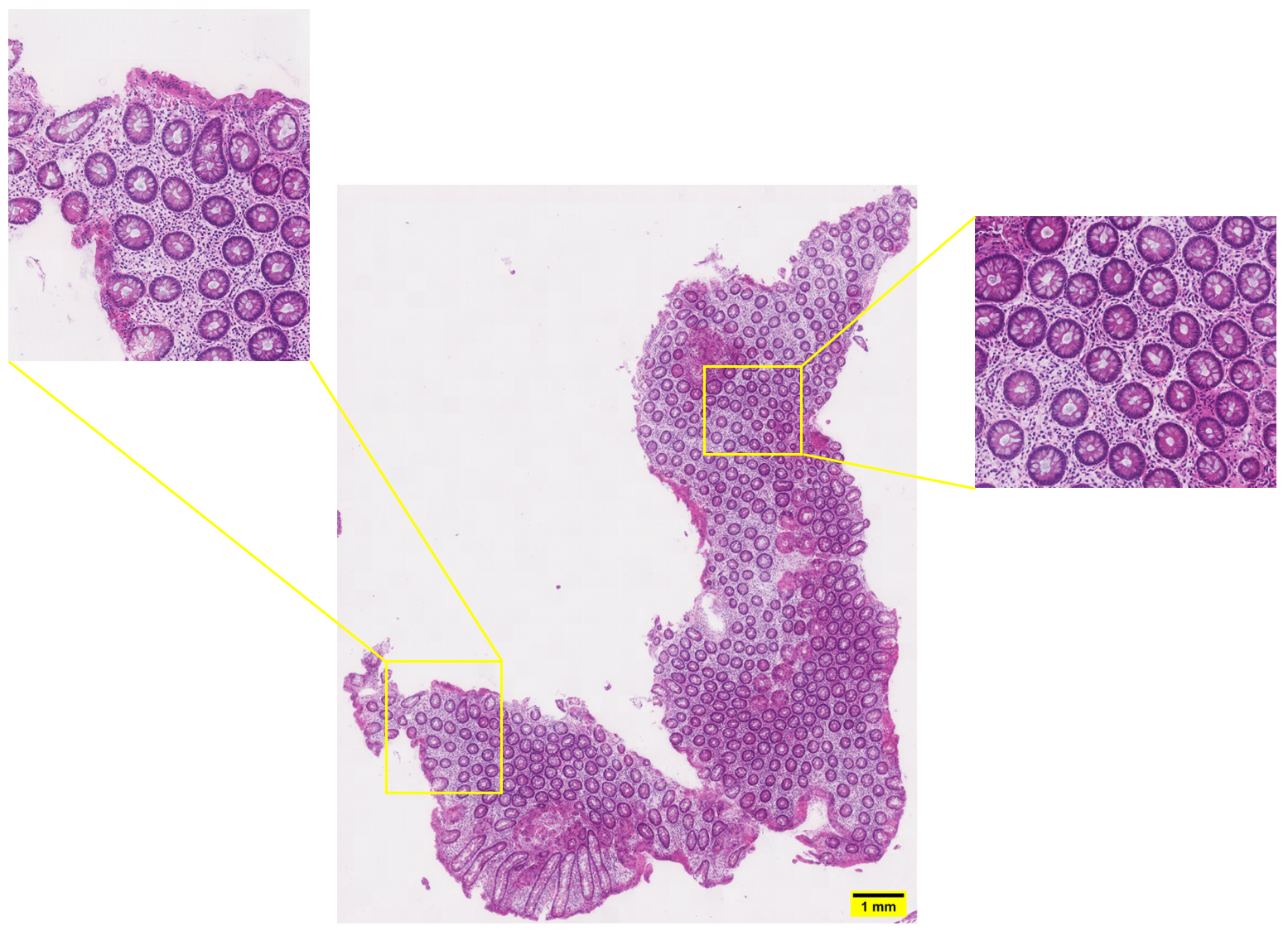}
\caption{Sample of generated image of size $7802\times9949$ Pixels}
\end{figure*}

\begin{figure*}[!hbpt]
\centering
\includegraphics[width=320pt,height=320pt]{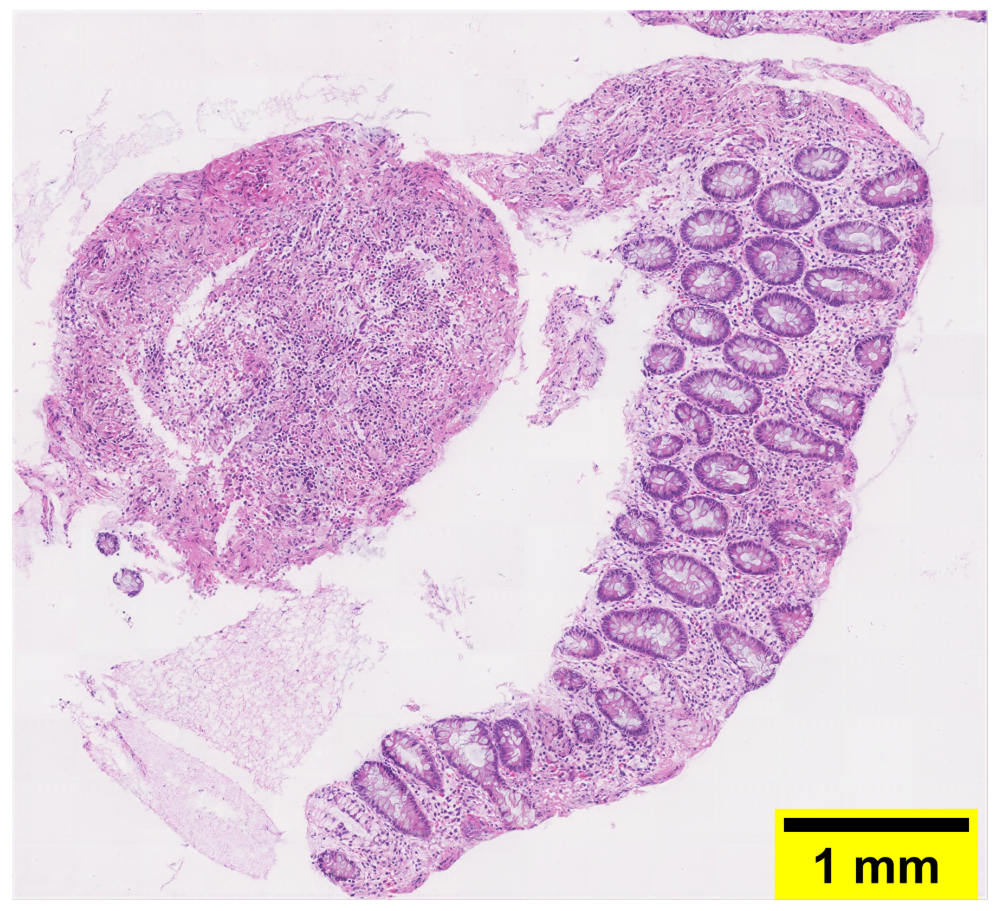}
\caption{Sample of generated image of size $3601\times3285$ Pixels}
\end{figure*}

\end{document}